\documentclass{achemso}
\usepackage{chemformula} % Formula subscripts using \ch{}
\usepackage[T1]{fontenc} % Use modern font encodings
\usepackage{amsmath} % Provides enhanced math symbols
\usepackage{amssymb} % Provides additional symbols
\usepackage{adjustbox}
\usepackage{rotating}
\usepackage{booktabs}
\usepackage{lscape}
\usepackage{graphicx}
\usepackage{longtable}
\usepackage{tabularx}
\usepackage{xcolor}
\usepackage{colortbl}
\author{Srilagna Sahoo}
\affiliation[Unknown University]{Department of Electrical Engineering, Indian Institute of Technology Bombay, Mumbai 400076, India }

\author{Abin Varghese}
\affiliation[Unknown University]{Department of Engineering, King’s College London, London WC2R 2LS, UK }

\author{Aniket Sadashiva}
\author{Mayank Goyal}
\author{Jayatika Sakhuja}
\author{Debanjan Bhowmik}
\author{Saurabh Lodha}

\email{slodha@ee.iitb.ac.in}

\affiliation[Unknown University]
{Department of Electrical Engineering, Indian Institute of Technology Bombay, Mumbai 400076, India }

\alsoaffiliation{https://orcid.org/0000-0002-0690-3169}

\title
  {Vertically Integrated Dual-memtransistor Enabled Reconfigurable Heterosynaptic Sensorimotor Networks and In-memory Neuromorphic Computing} 
\begin{document}

\setcitestyle{square} % Use square brackets in in-text citations

\makeatletter
% Use square brackets in bibliography
\renewcommand*{\@biblabel}[1]{[#1]}
\makeatother
%%%%%%%%%%%%%%%%%%%%%%%%%%%%% Keywords %%%%%%%%%%%%%%%%%%%%%%%%%%%%%%%%%%%
\section{Keywords}
2D materials, vdW, Heterostructure, Multiterminal, Ferroelectric, In$_2$Se$_3$, MoS$_2$, Homosynaptic and Heterosynaptic plasticity, Synaptic interactions, Synaptic cooperation, Synaptic competition, Logic gates

%%%%%%%%%%%%%%%%%%%%%%%%%%%%%%%%%%%%%%%%%%%%%%%%%%%%%%%%%%%%%%%%%%%%%
%% Abstract
%%%%%%%%%%%%%%%%%%%%%%%%%%%%%%%%%%%%%%%%%%%%%%%%%%%%%%%%%%%%%%%%%%%%%
\begin{abstract}
Neuromorphic in-memory computing requires area-efficient architecture for seamless and low latency parallel processing of large volumes of data. Here, we report a compact, vertically integrated/stratified field-effect transistor (VSFET) consisting of a 2D non-ferroelectric MoS$_2$ FET channel stacked on a 2D ferroelectric In$_2$Se$_3$ FET channel. Electrostatic coupling between the ferroelectric and non-ferroelectric semiconducting channels results in hysteretic transfer and output characteristics of both FETs. The gate-controlled MoS$_2$ memtransistor is shown to emulate homosynaptic plasticity behavior with low nonlinearity, low epoch, and high accuracy supervised (ANN - artificial neural network) and unsupervised (SNN - spiking neural network) on-chip learning. Further, simultaneous measurements of the MoS$_2$ and In$_2$Se$_3$ transistor synapses help realize complex heterosynaptic cooperation and competition behaviors. These are shown to mimic advanced sensorimotor neural network-controlled gill withdrawal reflex sensitization and habituation of a sea mollusk (Aplysia) with ultra-low power consumption. Finally, we show logic reconfigurability of the VSFET to realize Boolean gates thereby adding significant design flexibility for advanced computing technologies.   
\end{abstract}

\section{Introduction}
In today's data-intensive world, research in future computing paradigms is pursuing alternate paths compared to the traditional scaling of complementary metal oxide semiconductor (CMOS) transistor technology. Applications such as artificial intelligence, challenged by the von Neumann bottleneck, demand high performance as well as energy- and area-efficient computing devices with logic-in-memory functionality. This can be addressed by brain-inspired neuromorphic computing devices that offer in-memory computing with parallel processing capability along with low energy consumption and latency.\cite{thakar2023ultra} Their fundamental components are neurons (responsible for processing) that are connected to each other by synapses (responsible for storage) to form a neural network (NN). However, the challenges in realizing neuromorphic devices lie in material limitations that hinder the effective integration of memory and logic and, secondly, architecture that is both energy- and area-efficient. 

Ferroelectric field effect transistors (Fe-FETs) offer one such option of integrating logic and memory in a single, area-efficient device.\cite{ajayan2023ferroelectric} Conventional Fe-FETs involve stacking the semiconducting channel layer with an out-of-plane ferroelectric dielectric or polymer as gate dielectric to demonstrate a modulated memtransistor for synapses and logic circuits.\cite{lipatov2015optoelectrical, tian2019robust, chen2020van, ram2023reconfigurable} However, conventional bulk ferroelectric dielectrics (BaTiO$_3$, Pb(Zr, Ti)O$_3$, BiFeO$_3$) suffer from degradation of ferroelectric properties with thickness scaling, charge trapping, and poor interfaces.\cite{wang20232d}
In recent years, two-dimensional (2D) van der Waals (vdW) ferroelectrics have drawn attention because of their room temperature atomic level ferroelectricity, a dangling bond-free surface and high-quality heterointerfaces due to the absence of lattice mismatch with other materials. \cite{baek2022ferroelectric, liu2016room, wang2023towards, ghosh2022polarity, varghese2020near} CuInP$_2$S$_6$ (CIPS) is one such 2D vdW ferroelectric dielectric that shows out-of-plane (OOP) ferroelectricity in ultra-thin flakes.\cite{belianinov2015cuinp2s6,liu2016room,baek2022ferroelectric} Unlike conventional insulating ferroelectrics, 2D vdW ferroelectrics can also be semiconducting in nature. This offers an additional degree of freedom in the design of ferroelectric semiconductor (FeS) FETs (FeS-FETs) for logic-in-memory functionality.\cite{si2019ferroelectric,liao2023van} Ferroelectric 2D vdW semiconductors such as SnSe,\cite{chang2020microscopic} and SnS \cite{higashitarumizu2020purely, bao2019gate} show in-plane (IP) ferroelectric polarization.\cite{wang2023towards} In$_2$Se$_3$ on the other hand shows both IP and OOP ferroelectricity, which makes it suitable for multi-directional conductivity control device architectures.\cite{xiao2018intrinsic, cui2018intercorrelated} Several memory device and neural network designs using In$_2$Se$_3$ as a semiconductor channel or a ferroelectric gate have been reported in conjunction with CIPS\cite{baek2022ferroelectric}, MoS$_2$\cite{park2023laterally}, and hBN\cite{wang2021two,kang2024photo}. However, reported structures are either area-inefficient (lateral or multi-device architecture), and/or they do not simultaneously exploit their semiconducting and correlated IP and OOP ferroelectric polarization properties.
 
In this work, we demonstrate a compact, vertically stacked heterostructure FET where the 2D vdW semiconductor (S) MoS$_2$ is stacked upon ferroelectric semiconductor (FeS) In$_2$Se$_3$  on gate insulator (I) hBN aligned over the metal (M) gate to form a MIFeSS structure in a top-down approach. Lateral electrodes are placed on both semiconductors such that they can be readily accessed simultaneously. The FeS In$_2$Se$_3$ is stacked on top of the gate electrode and beneath the MoS$_2$ to gain maximum control on its polarization and mobile charge, while its OOP polarization can be leveraged to tune MoS$_2$ conduction. This densely packed vertical stratified field-effect transistor (VSFET) architecture utilizing a non-FeS/FeS heterostructure harnesses not only the semiconducting properties of each layer but also the interface FeS-S coupling between them.

This work consists of three parts. The first part utilizes the electrostatic coupling between MoS$_2$ (S) and In$_2$Se$_3$ (FeS) to demonstrate hysteretic behavior in transfer (drain current vs. gate voltage: $I_D$ - $V_G$) and output (drain current vs. drain voltage: $I_D$ - $V_D$) characteristics of the MoS$_2$ FET. The gate-modulated MoS$_2$ memtransistor is shown to emulate biological input-specific Hebbian homosynaptic plasticity behavior with low nonlinearity, and high accuracy supervised (ANN - artificial neural network) and unsupervised (SNN - spiking neural network) on-chip learning. In the second part, we demonstrate advanced biomimetic heterosynaptic cooperation and competition between the two simultaneously measured MoS$_2$ and In$_2$Se$_3$ memtransistors representing two distinct but coupled synapses. This heterosynaptic behavior is shown to replicate the highly developed sensorimotor NN governing the gill reflexes of a sea mollusk. Finally, the realization of two-input Boolean logic gates comprehensively shows the area-efficient, logic, and multi-memory reconfigurability of the VSFET architecture.

%%%%%%%%%%%%%%%%%%%%%%%%% Results %%%%%%%%%%%%%%%%%%%%%%%%%%%%%%%%%%%%
\section{Results}
\begin{figure}[H]	
{\includegraphics[width=1\textwidth]{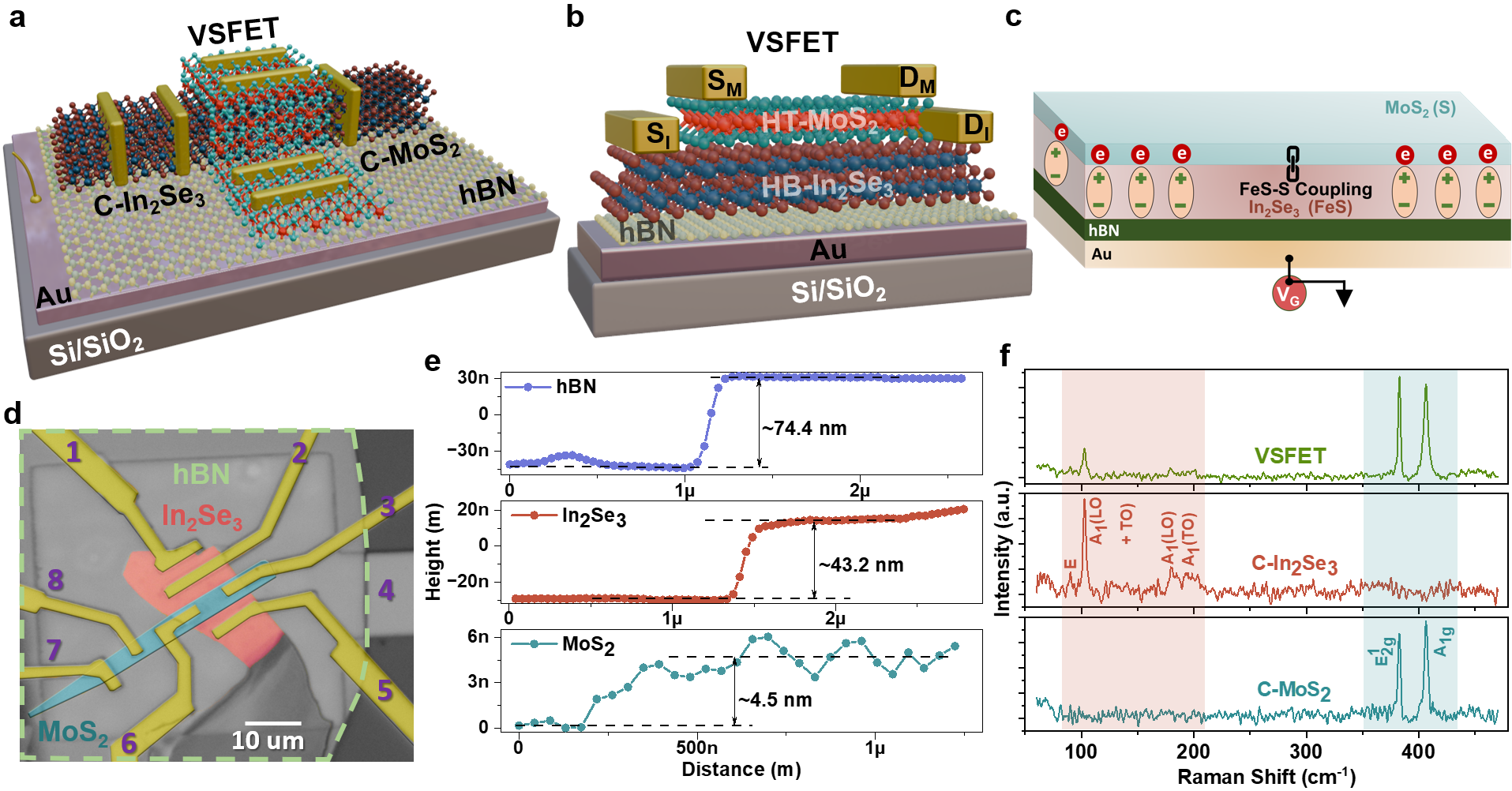}}
\caption{Device structure and Raman spectra. (a) 3D schematic of  MoS$_2$ and In$_2$Se$_3$ vertical stratified field-effect transistor (VSFET) with hBN as the gate dielectric. (b) 2D cross-section of MoS$_2$/In$_2$Se$_3$ VSFET. (c) Pictorial representation of 2D ferroelectric and non-ferroelectric semiconductor coupling that modulates HT-MoS$_2$ electron current. (d) False color optical microscope image of the fabricated vdW heterostructure VSFET on Au/SiO$_2$/p$^{+}$ Si substrate. Dashed line indicates the boundary of the hBN flake. (e) Thickness profiles of hBN, In$_2$Se$_3$, and MoS$_2$ characterized by atomic force microscopy. (f) Raman spectra acquired from C-MoS$_2$, C-In$_2$Se$_3$, and the MoS$_2$/In$_2$Se$_3$ heterojunction regions.}
\label{schematic}
\end{figure}

\textbf{Device Architecture:}
A three-dimensional (3D) schematic of the VSFET stack consisting of Au (bottom gate)/hBN (dielectric)/n-type In$_2$Se$_3$ (FeS)/n-type MoS$_2$ (top S) is shown in Figure \ref{schematic}a. 2D cross-section of the MoS$_2$/In$_2$Se$_3$ VSFET that shows heterostructure stacking of two individual FET channels (top MoS$_2$: HT-MoS$_2$ and bottom In$_2$Se$_3$: HB-In$_2$Se$_3$) is depicted in Figure \ref{schematic}b, enabling 2D vdW FeS-S coupling that can modulate the HT-MoS$_2$ electron conduction (Figure \ref{schematic}c). The bottom metal gate was patterned on a 285 nm SiO$_2$/Si substrate using electron beam (e-beam) lithography and metal deposition (4 nm Ti/30 nm Au) by sputtering. A thick (74.4 nm) hBN flake was then transferred onto the Ti/Au gate electrode. Next, a thick (43.2 nm) In$_2$Se$_3$ flake was transferred onto the hBN and metal gate overlap area. Finally, a thin (4.5 nm) MoS$_2$ flake was transferred on top of the In$_2$Se$_3$ flake transversely such that some part of MoS$_2$ forms a heterojunction with the In$_2$Se$_3$ flake, and some part gets transferred directly onto the hBN flake to form a control MoS$_2$ (C-MoS$_2$) FET (Figure S1a) that works as a control device without In$_2$Se$_3$. Likewise, the exposed part of In$_2$Se$_3$ on the hBN flake forms the control In$_2$Se$_3$ (C-In$_2$Se$_3$) FET that serves as the other control device without MoS$_2$ (Figure S1a). Following all the flake transfers, the source/drain (S/D) electrodes were e-beam patterned and metallized (4 nm Ti/100 nm Au) by sputtering. A false color optical micrograph of the fabricated device is shown in Figure \ref{schematic}d. Thicknesses of the hBN, In$_2$Se$_3$, and MoS$_2$ layers characterized by atomic force microscopy (AFM) scans are reported in Figure \ref{schematic}e.

Raman scattering peaks of C-MoS$_2$ were observed at 382.4 cm$^{-1}$ and 406.3 cm$^{-1}$ corresponding to E$_{2g}$ and A$_{1g}$ phonon modes respectively (Figure \ref{schematic}f).\cite{kaushik2014schottky, jawa2022wavelength} A difference of 24 cm$^{-1}$ between them indicates $\geq$ 4-layer thickness of MoS$_2$, but confirmed to be $\sim$6-layer thick based on AFM height profile (Figure \ref{schematic}e).\cite{yuan2017interfacial, lee2010anomalous}  The four Raman peaks observed at 90.1 cm$^{-1}$, 102.6 cm$^{-1}$, 180.4 cm$^{-1}$, and 193.2 cm$^{-1}$ indicate the typical E, A$_1$(LO+TO), A$_1$(LO), and A$_1$(TO) phonon modes of $\alpha$-phase In$_2$Se$_3$ respectively (Figure \ref{schematic}f).\cite{liu2022optoelectronic, dutta2021cross} The Raman peak at 90.1 cm$^{-1}$ indicates the hexagonal symmetry of $\alpha$-In$_2$Se$_3$.\cite{xue2018multidirection} The presence of  A$_1$(LO) and A$_1$(TO) phonon modes due to LO-TO splitting indicates the lack of inversion symmetry, making $\alpha$-In$_2$Se$_3$ ferroelectric.\cite{dutta2021cross} Signature peaks of both materials are also present in the VSFET overlap region. $\alpha$-In$_2$Se$_3$ exhibits correlated IP and OOP polarization \cite{cui2018intercorrelated}, wherein OOP polarization is due to a breaking of centrosymmetry due to the asymmetric position of the middle Se layer in the 5-layer (Se-In-Se-In-Se) structure of In$_2$Se$_3$. \cite{xiao2018intrinsic} Polarization switching in the In$_2$Se$_3$ layer was characterized by piezoresponse force microscopy (PFM) (Figures S1b-c). A butterfly-like amplitude vs. voltage loop and a phase shift of \(180^\circ\) were obtained, indicating ferroelectric switching in In$_2$Se$_3$.\cite{wang2020exploring, varghese2024electrically}

%%%%%%%%%%%%figure1

\begin{figure} [H]	
{\includegraphics[width=1\textwidth]{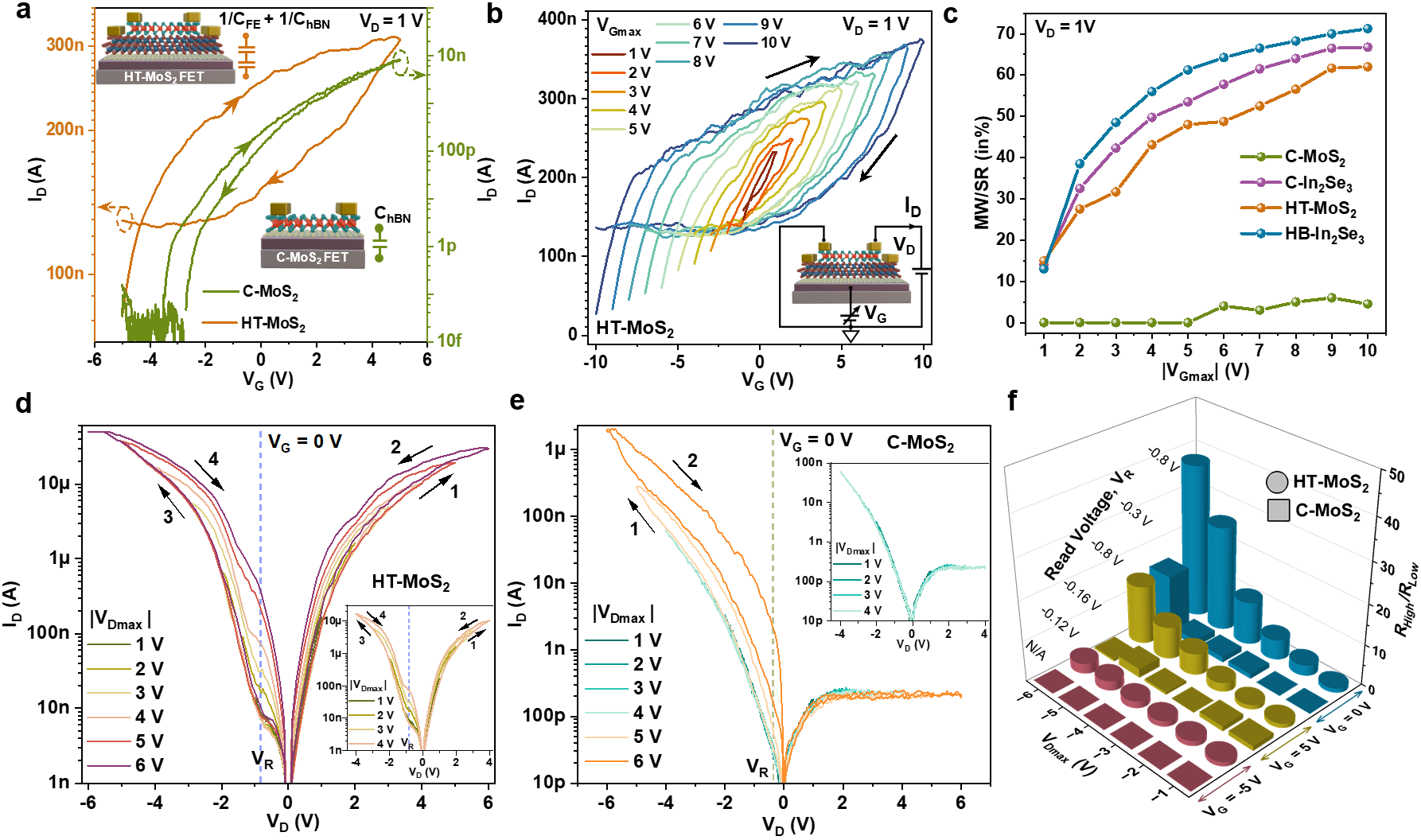}}
\caption{DC electrical characteristics of VSFET. (a) Comparison of transfer characteristics between C-MoS$_2$ and HT-MoS$_2$ FETs at V$_D$ = 1 V showing notable hysteretic behavior in HT-MoS$_2$ in contrast to C-MoS$_2$. (b) $I_D$ - $V_G$  characteristics of HT-MoS$_2$ FET under different gate bias sweeps. (c) MW/SR ratios of C-MoS$_2$, C-In$_2$Se$_3$, HT-MoS$_2$, and HB-In$_2$Se$_3$ FETs as a function of maximum gate bias sweep range. (d) Pinched output hysteresis loops ($I_D$ - $V_D$) of HT-MoS$_2$ under different drain bias sweep ranges. The inset shows pinched output hysteresis loops for |$V_{Dmax}$| = 1 V $\rightarrow$ 4 V, indicating the gradual opening of the resistive window even at lower voltages. However, (e) output curves of  C-MoS$_2$ FET show no such resistive window opening for |$V_{Dmax}$| = 1 V $\rightarrow$ 4 V (inset). For $V_{Dmax}$ $\geq  $ -5 V, a non-gradual resistive window opening can be seen likely due to higher $V_{D}$. (f) The resistance ratio versus $V_{Dmax}$ for C-MoS$_2$ and HT-MoS$_2$ FETs for three different gate voltages (0 V, 5 V, -5 V).
}
\label{electrical}
\end{figure}

\textbf{Electrical Characterization:}  Transfer characteristics of both the MoS$_2$ (C-MoS$_2$ and HT-MoS$_2$) FETs show n-type behavior and clockwise hysteresis with gate bias sweeps from -5 V $\rightleftharpoons$ 5 V under $V_D$ = 1 V (Figure \ref{electrical}a). However, HT-MoS$_2$ FET shows significantly larger hysteresis because of the underlying ferroelectric In$_2$Se$_3$, compared to the C-MoS$_2$ FET. The C-MoS$_2$ FET shows an on-/off- current ($I_{on}$/$I_{off}$) ratio of $10^6$, whereas HT-MoS$_2$ FET shows a 3$\times$ ratio. The low ratio is due to the extra ferroelectric capacitance ($C_{FE}$) from In$_2$Se$_3$ in series with the dielectric capacitance ($C_{hBN}$) in the HT-MoS$_2$ FET, resulting in poor gate modulation. Transfer curves with 0.5 V drain bias for HT-MoS$_2$ and C-MoS$_2$ FETs are shown in Figure S3a. 

Further, to investigate the origin of hysteresis in the HT-MoS$_2$ FET, $I_D$ was measured with maximum gate sweep voltage (|$V_{Gmax}$|) increasing from 1 V to 10 V (Figure \ref{electrical}b). The hysteresis loop memory window (MW) expansion with increasing |$V_{Gmax}$| is due to gate field-controlled gradual increase in ferroelectric polarization charge of the underlying In$_2$Se$_3$ layer. Here, MW refers to the maximum hysteresis window seen at a constant value of drain current. This indicates effective modulation of carrier transport in the MoS$_2$ channel due to ferroelectric coupling with In$_2$Se$_3$.\cite{wang2021two} In contrast, increasing |$V_{Gmax}$| does not increase the hysteresis window in the $I_D$ - $V_G$ loops for the C-MoS$_2$ FET, as shown in Figure S2a. Progressive increase of the hysteresis window in the $I_D$ - $V_G$ loops for C-In$_2$Se$_3$ and HB-In$_2$Se$_3$ FETs is shown in Figures S2b-c. Similar progressive increase in the hysteretic memory window at a lower 0.5 V drain bias is shown in Figures S3b-e for C-MoS$_2$, HT-MoS$_2$, C-In$_2$Se$_3$, and HB-In$_2$Se$_3$, respectively. The MW-to-sweep range (SR) ratio (MW/SR) of all four FETs versus |$V_{Gmax}$| is shown in Figure \ref{electrical}c, indicating a monotonic increase in the  MW/SR ratio for HT-MoS$_2$ with |$V_{Gmax}$|, like the In$_2$Se$_3$ FETs (C- and HB-), and unlike the C-MoS$_2$ FET. The MW/SR ratio measurements were repeated on two additional HT-MoS$_2$ FETs with similar results (Figure S2d). The MW/SR data of all four FETs for $V_D$ = 0.5 V is shown in Figure S3f. Gate tunable output ($I_D$ - $V_D$) characteristics of C-MoS$_2$, HT-MoS$_2$, C-In$_2$Se$_3$, and HB-In$_2$Se$_3$ FETs are shown in Figure S4.

While the OOP polarization charge of HB-In$_2$Se$_3$ modulates HT-MoS$_2$ transfer characteristics, the IP polarization charge of HB-In$_2$Se$_3$ also modulates HT-MoS$_2$ output characteristics at 0 V gate bias (no OOP field, Figure \ref{electrical}d). Sweeping the drain bias along 0 V $\rightarrow$  +$V_{Dmax}$ $\rightarrow$ 0 V and then 0 V $\rightarrow$ -$V_{Dmax}$ $\rightarrow$ 0 V, the output curve shows a hysteresis loop (directions are marked in black arrows) pinched at 0 V. These pinched hysteresis loops (PHLs) open up gradually from a closed state (|$V_{Dmax}$| = 1 V) with increasing |$V_{Dmax}$|. This leads to the appearance of two distinct resistive states $R_{High}$ and $R_{Low}$ for higher |$V_{Dmax}$| values, where $R$ = $V_{D}$/$I_{D}$. The resistive switching behavior could be attributed to IP (lateral) ferroelectric domain propagation in HB-In$_2$Se$_3$ that affects the lateral transport characteristics in HT-MoS$_2$ through Schottky barrier modulation at the contacts. The resistance window opens up after the IP coercive voltage ($V_{cip}$) is reached. 1 V < $V_{cip}$ < 2 V and beyond |$V_{Dmax}$| $\geq  $ 2 V, the resistance ratio (RR): $R_{High}$/$R_{Low}$ at a read drain bias voltage $V_R$, increases till $V_{Dmax}$ = -5 V and saturates thereafter due to the saturation of In$_2$Se$_3$ IP polarization.\cite{xue2021giant} The asymmetry observed in PHLs for +$V_{Dmax}$ and -$V_{Dmax}$  is likely due to the asymmetry in contact areas. The inset of Figure \ref{electrical}d shows pinched output hysteresis loops for |$V_{Dmax}$| = 1 V $\rightarrow$ 4 V, indicating the gradual opening of the resistive window even at lower voltage ranges. However, the output curve of C-MoS$_2$ FET (Figure \ref{electrical}e) shows no such resistive window opening for |$V_{Dmax}$| = 1 V → 4 V (inset of Figure \ref{electrical}e). For $V_{Dmax}$ $\geq$ -5 V, a non-gradual hysteresis window opening can be seen, most likely due to a high electric field that can introduce defects (field-induced trap generation), resulting in a threshold voltage shift. The RRs for negative drain bias, shown in Figure \ref{electrical}f, are higher compared to positive drain bias. RRs\textbf{, }extracted at a fixed drain bias $V_R$, for three different gate biases versus |$V_{Dmax}$|, are shown for C-MoS$_2$ and HT-MoS$_2$ FETs.  HT-MoS$_2$ has a higher RR than C-MoS$_2$ due to its HB-In$_2$Se$_3$ IP ferroelectric polarization modulated conductance. In summary, the intercorrelated IP-OOP polarization of In$_2$Se$_3$ can effectively tune both the trans- and output-conductance of HT-MoS$_2$.

%%%%%%%%%%%%figure2

\begin{figure} [H]	
{\includegraphics[width=1\textwidth]{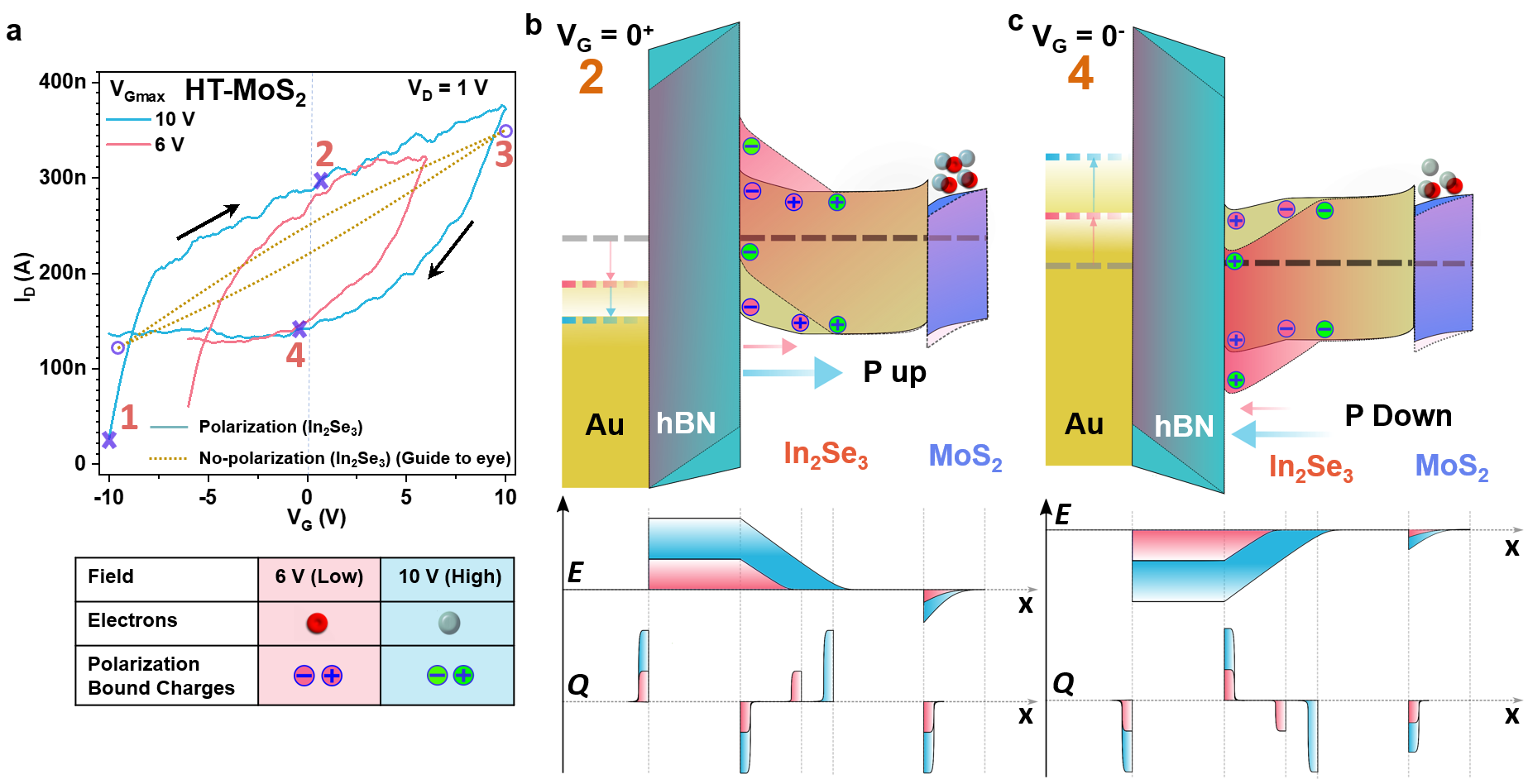}}
\caption{Device physics for |$V_{Gmax}$| dependent $I_D$ - $V_G$ hysteresis in HT-MoS$_2$ FET. (a) The transfer curve of HT-MoS$_2$ FET on the left has four marked points, 1, 2, 3, and 4, representing four operating points of the device in a clockwise direction for |$V_{Gmax}$| = 10 V. Energy band diagrams for (b) point 2 and (c) point 4 illustrate the change in electrostatics, spatial electric field, and charge distribution corresponding to a change in |$V_{Gmax}$| = 6 V (red) $\rightarrow$  10 V (blue). Point 2 (point 4) indicates a low positive (negative) gate field induced weak upward (downward) partial In$_2$Se$_3$ polarization and stronger (weaker) MoS$_2$ accumulation than without polarization. The higher (point 2) and lower (point 4) MoS$_2$ currents vs the no-polarization case open up a hysteresis memory window.}
\label{EBD}
\end{figure}

\textbf{Device Physics:} The clockwise hysteretic behavior in the transfer characteristics of HT-MoS$_2$ (Figure \ref{electrical}a) can be explained using the local gate field and ferroelectric polarization in In$_2$Se$_3$ as shown in Figure \ref{EBD} through energy band diagrams (EBDs) of the Au/hBN/In$_2$Se$_3$/MoS$_2$ stack at four representative $V_G$ conditions. EBDs of each individual material used in the heterostructure before contact \cite{kaushik2014schottky,jawa2021electrically, jawa2022wavelength, mu2024homo, lyu2020thickness, elias2019direct, cassabois2016hexagonal} and at equilibrium after contact are described in supporting information in Figure S5a-b. Equilibrium EBD (Figure S5b) indicates an accumulated MoS$_2$ interface with In$_2$Se$_3$ consistent with a conducting MoS$_2$ channel at $V_G$ = 0 V (transfer curve in Figure \ref{EBD}a). 

%%%%%%%%%%%%%point 1
At -$V_{Gmax}$ of -10 V (point 1 (Figure S5c) on $I_D$ - $V_G$ curve), the electric field across the 43 nm thick In$_2$Se$_3$ layer is higher than its coercive field, resulting in a fully polarized In$_2$Se$_3$ layer with positive polarization charge at the hBN/In$_2$Se$_3$ interface and negative fixed polarization charge at the In$_2$Se$_3$/MoS$_2$ interface. The series combination of hBN and In$_2$Se$_3$ capacitances leads to a lower effective capacitance and poor modulation of the drain current, which stays high, and HT-MoS$_2$ does not turn off entirely. However, the negative polarization charge near the In$_2$Se$_3$/MoS$_2$ interface reduces electron concentration in the MoS$_2$ channel and the off-current is lower than what it would have been without the OOP polarization in In$_2$Se$_3$.

%%%%%%%%%%%%%point 3
Likewise, at +$V_{Gmax}$ of  +10 V (point 3 (Figure S5c) on $I_D$ - $V_G$ curve), the polarization field in the In$_2$Se$_3$ layer flips such that we have negative charge at the hBN/In$_2$Se$_3$ interface and positive at the In$_2$Se$_3$/MoS$_2$ interface. The positive polarization charge near the In$_2$Se$_3$/MoS$_2$ interface increases electron concentration in the MoS$_2$ channel and the on-current is higher than what it would have been without the OOP polarization in In$_2$Se$_3$.

%%%%%%%%%%%%%point 2 & 4
However, the origin of the hysteresis is due to the difference in the electrostatics at points 2 and 4 (EBDs in Figure \ref{EBD}b-c) of the $I_D$ - $V_G$ curve. Here, EBDs for point 2 and point 4 illustrate the change in electrostatics, spatial electric field, and charge distribution corresponding to a change in |$V_{Gmax}$| from 6 V (low field, red) to 10 V (high field, blue). At point 2 ($V_G$ = \( 0^+ \), forward $V_G$ sweep) for high |$V_{Gmax}$| = 10 V, the In$_2$Se$_3$ layer is partially (upward) polarized but with a more accumulated MoS$_2$ channel (higher current than without polarization), similar to the electrostatics at point 3 (Figure S5c). Conversely, at point 4 ($V_G$ = \( 0^- \), reverse $V_G$ sweep) for high |$V_{Gmax}$| = 10 V, the In$_2$Se$_3$ is partially (downward) polarized but with a less accumulated MoS$_2$ channel (lower current than without polarization), similar to the electrostatics at point 1 (Figure S5c). The higher on-current and lower off-currents at points 2 and 4, compared to the no-polarization case, open up a hysteresis memory window.

Since the partial polarization strengths at points 2 and 4 depend on the maximum gate field values (|$V_{Gmax}$|) and the polarization strengths at points 1 and 3, the MW is a function of applied |$V_{Gmax}$|. The strengths of the polarization field, polarization bound charge and their position change with an increase in |$V_{Gmax}$| from 6 V to 10 V, thereby changing the electron charge density (current) in the MoS$_2$ channel at point 2 and 4. The resulting memory window expansion with increasing |$V_{Gmax}$| can be seen in the transfer curves (Figure \ref{EBD}).

%%%%%%%%%%%%figure3

\begin{figure} [H]	
{\includegraphics[width=1\textwidth]{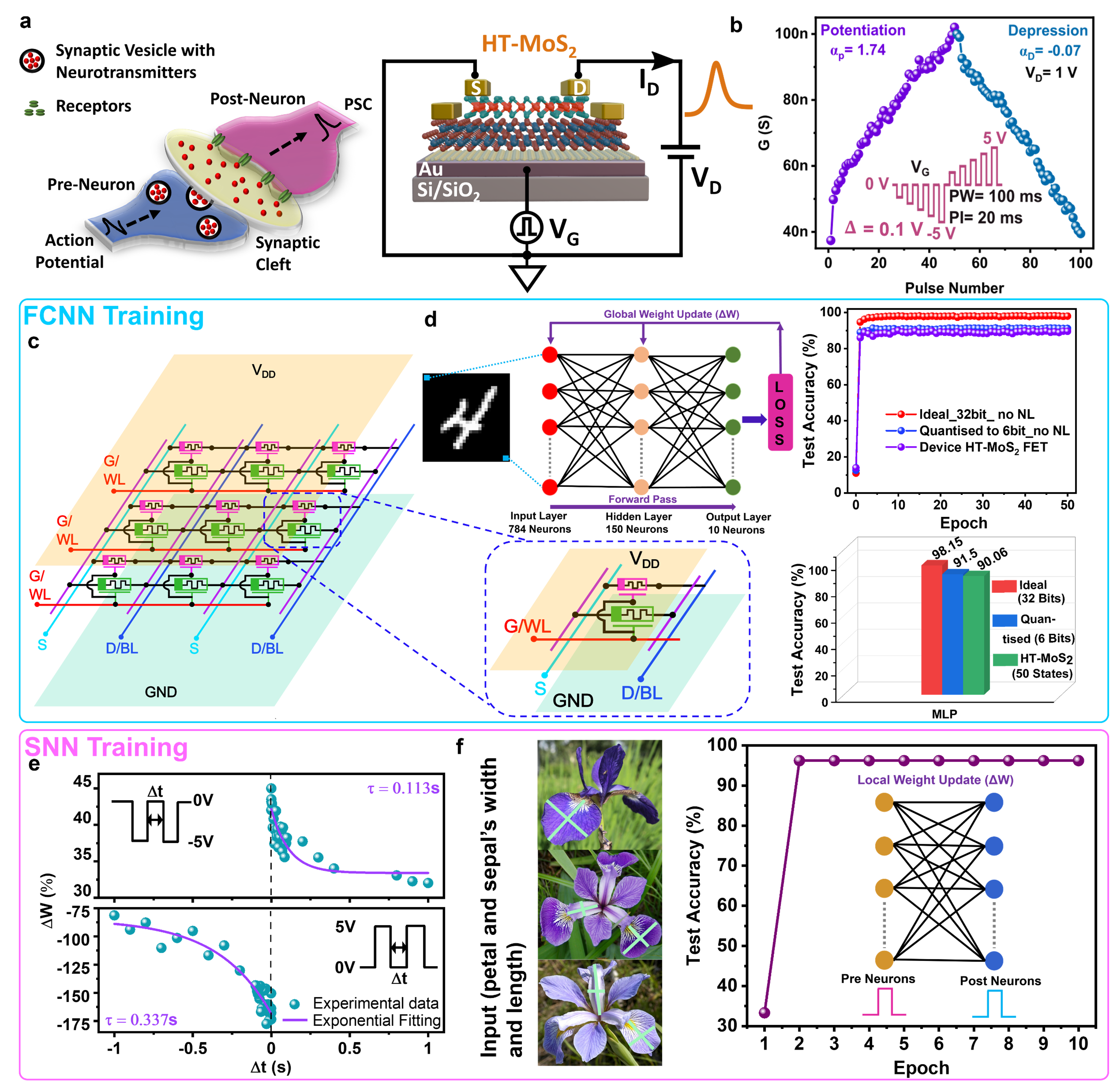}}
\caption{Neuromorphic in-memory computing using HT-MoS$_2$ synapse. (a) Biological synaptic junction equivalent electrical circuit for HT-MoS$_2$ FET under gate and source/drain biases. (b) Measured potentiation and depression characteristics of HT-MoS$_2$ synaptic FET. (c) 3D schematic of HT-MoS$_2$ FET integration in standard crossbar structure. (d) Fully connected neural network (FCNN) of non-spiking type trained to classify MNIST handwritten digits using a back-propagation algorithm with weight update limited to one potentiation/depression step at any iteration. Each input neuron corresponds to a unique pixel in the image. On-chip learning accuracy numbers (in data plot and bar chart) are shown for: synaptic weights with ideal 32-bit precision, with 6-bit precision without non-linearity, and those following the experimentally obtained potentiation and depression curves as shown in (b). (e) Measured STDP behavior of HT-MoS$_2$ synaptic FET. (f) Training accuracy of a spiking neural network (SNN) on Fisher's iris dataset, using experimental STDP data of (e) for weight update during training. \cite{iris_setosa_image,iris_versicolor_image,iris_virginica_image}}
\label{NN}
\end{figure}

\textbf{Homosynaptic Plasticity and Neuromorphic In-Memory Computing:} Conductance modulation in HT-MoS$_2$ FET through gate-controlled ferroelectric polarization of the HB-In$_2$Se$_3$ layer can be leveraged for synapse functionality. Figure \ref{NN}a illustrates a biological synapse and a corresponding electrical biasing scheme for HT-MoS$_2$ FET where the gate terminal serves as the presynaptic axon end, the source terminal serves as the starting point of the synaptic cleft, and the drain terminal mimics the postsynaptic end that collects the postsynaptic current. Figure S6a (S6b) shows the excitatory (inhibitory) post-synaptic current (PSC) response for increasing negative (positive) amplitude pulses applied at the gate, due to the modulation in partial polarization consistent with the multi-|$V_{Gmax}$| sweeps.\cite{wang2020exploring} Figures S6c and d show the same for different pulse widths.  Figure S6e shows the paired-pulse facilitation data for varying pulse separation times, as illustrated in Figure S6f. 

Potentiation (P, increase in synaptic strength) and depression (D, decrease in synaptic strength) are two key synaptic plasticity behaviors that help in learning and developing varied functionalities in neural circuits. Figure \ref{NN}b shows the P/D behavior of the HT-MoS$_2$ synapse while applying 50 (50) incremental pulses (width = 100 ms, interval = 20 ms) in the negative (positive) direction from -0.1 V to -5 V (0.1 V to 5 V) with a step size of  0.1 V. During potentiation, the synapse conductance changed from 37.4 nS to 102 nS with non-linearity factor $\alpha$$_p$ = 1.74. Likewise, during the depression, it changed from 100 nS down to 39.5 nS over 50 pulses with a non-linearity factor $\alpha$$_d$ = -0.07. The different nonlinear rates for P and D result in an asymmetry factor ($\alpha$$_p$-$\alpha$$_d$) of 1.81 (see details in supporting information 7). Figure S8a shows multiple cycles of P/D data using this pulse scheme, whereas S8b and S8c show the P/D data for constant amplitude and varying pulse width schemes, respectively. The nonlinearity parameters for all the pulse schemes are tabulated in Table 1, supporting information 8.

Next, the P/D parameters extracted from Figure \ref{NN}b were used to model the hardware implementation of a NN circuit utilizing MoS$_2$/In$_2$Se$_3$ vdW VSFET arrays in crossbar architecture, as shown in Figure \ref{NN}c. A single cell of the crossbar matrix consists of two distinct memtransistors: 1) HT-MoS$_2$ FET at the top and 2) HB-In$_2$Se$_3$ FET at the bottom. Interestingly, the conductance levels of the two FETs are coupled owing to a common gate terminal. This coupling enables multiple additional functionalities, which will be discussed later. Here, the coupled gate terminal functions as the word line (WL), while the drain terminal of each HT-MoS$_2$ FET connects to the bit line (BL). The source terminal of each HT-MoS$_2$ FET is connected to the ground (GND), and $V_{DD}$ serves as the supply voltage. Supervised learning accuracy numbers using a non-spiking multilayer perceptron neural network (MLP), or fully connected neural network (FCNN), were evaluated for digit classification, as shown in Figure \ref{NN}d. The three-layer MLP contains 784 input neurons in the first layer, 150 hidden neurons in the second layer, and 10 output neurons in the third layer (Figure \ref{NN}d). A 28x28 pixel image from the MNIST dataset is shown in Figure \ref{NN}d; it was given as input for the classification task. Each image pixel corresponds to one input neuron in the first layer.

For the learning accuracy evaluation, a back-propagation algorithm was used to update the synaptic weights at any iteration, limited to 1-step of potentiation or depression, consistent with the pulsing scheme. This 1-step or 1-bit weight update makes the crossbar array and peripheral circuit design simpler. Tanh and softmax neuronal activation functions were used for the hidden and output layers, respectively.  Conductance non-linearity was considered as a non-ideal parameter during the MLP training. More details of the neural network design and the training algorithm can be found in reports by Kaushik et al. \cite{kaushik2020synapse}, and Yadav et al. \cite{yadav2023demonstration} 60,000 out of  70,000 MNIST dataset samples were used as the training data, and the remaining 10,000 samples were used for testing. Figure \ref{NN}d shows that test accuracy when each synapse is modeled after the device experimental data, including its nonlinearity and asymmetry (Figure \ref{NN}b), reaches 90.06 \% within 5 epochs. Test accuracy numbers for the network 32-bit ideal synaptic weights and 6-bit quantized synaptic weights are 98.15 \% and 91.5 \% respectively. The on-chip inference test accuracy data, where synaptic weight updates are based on all the three pulse schemes described earlier, is tabulated in Table 2, supporting information 9. The energy consumption per spike (E$_{per spike}$) is 1.38 nJ for a -5 V gate voltage spike during potentiation and 16.8 nJ for a 5 V gate voltage spike during depression. $E_{per spike}$ = $I_{PSC}$ * $V_D$ * $t_{spike}$, where $I_{PSC}$ denotes the maximum postsynaptic current and $t_{spike}$ denotes the pulse width of the presynaptic voltage spike.\cite{hwang2023bioinspired}

The HT-MoS$_2$ FET also exhibits spike time-dependent plasticity (STDP), \cite{crair2009long}  wherein synaptic strength depends upon the causal relationship between pre- and post-neuron firing. \cite{gutig2014spike} Synaptic efficacy gets potentiated (weakened) when the pre-neuron fires before (after) the post-neuron in a few ms temporal gap. The time interval $\Delta$t = $t_{post}$ - $t_{pre}$, between post- and pre-pulse occurrence, decides the weight update ($\Delta W$),  \begin{equation}
\Delta W (\%) = 100 \times \frac{G_{\text{after}} - G_{\text{before}}}{G_{\text{before}}}
\end{equation} where $G_{before}$, and $G_{after}$ denote the conductance values before and after one iteration of paired-pulse application separated by $\Delta$t. Figure \ref{NN}e shows the measured STDP data for the HT-MoS$_2$ synapse for positive and negative 5 V pulses separated by varying $\Delta$t. Exponential fits (supporting information 10) for decrease and increase in synaptic strength for positive and negative $\Delta$t yield time constants of 113 ms and 337 ms, respectively. Unsupervised learning tested on Fisher’s iris dataset using a spiking neural network (SNN) \cite{sahu2019spike} based on this STDP data yields a stable test accuracy of 96.2 \% after 2 epochs, as shown in Figure \ref{NN}f.

%%%%%%%%%%%%figure4

\begin{figure} [H]	
{\includegraphics[width=1\textwidth]{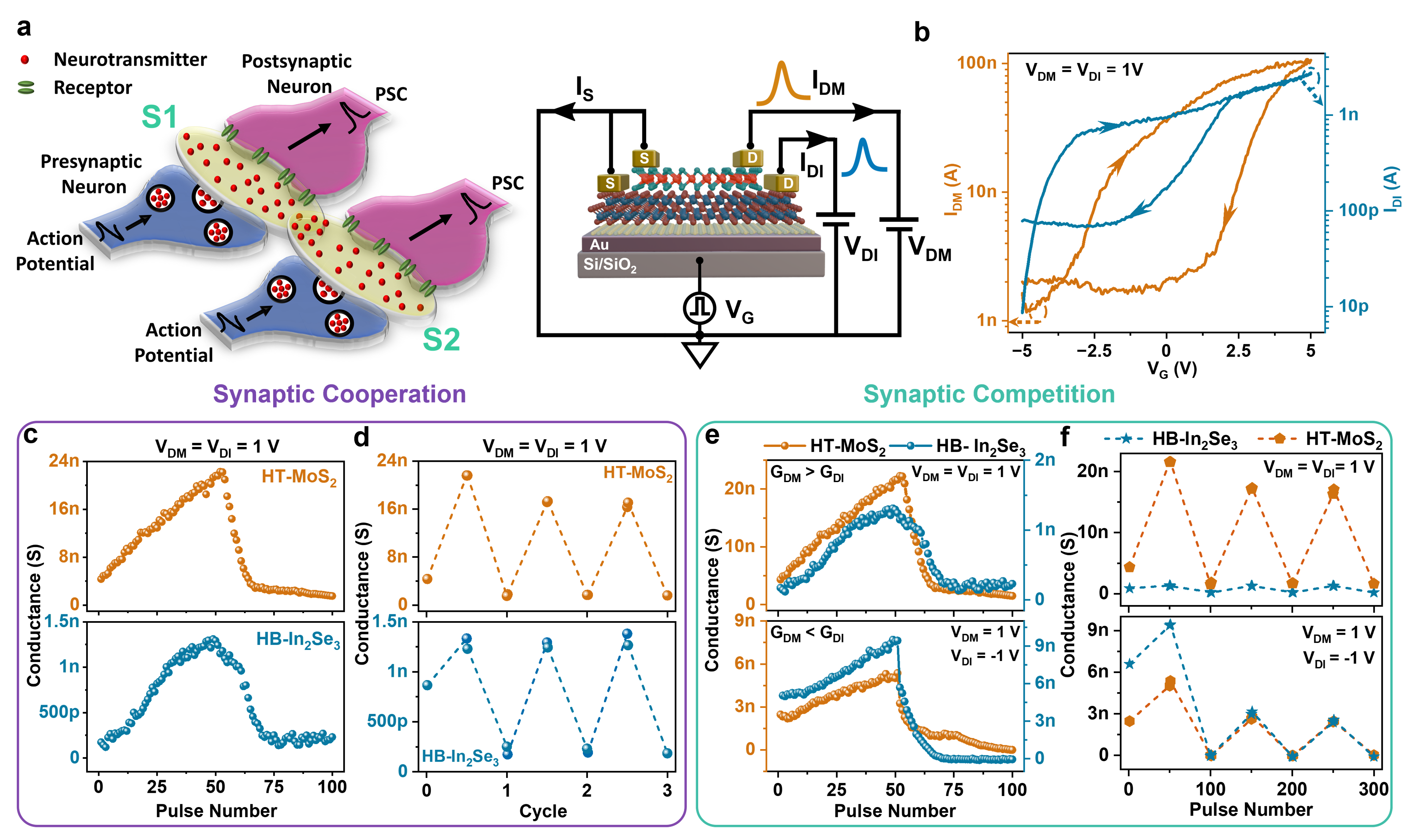}}
\caption{Heterosynaptic interactions utilizing VSFET. (a) Schematics showing two interacting biological synapses and their five-terminal equivalent electrical circuit representation using the VSFET. (b) Simultaneously extracted $I_D$ - $V_G$ curves of HT-MoS$_2$ and HB-In$_2$Se$_3$ FETs under the biasing scheme shown in (a). Synaptic cooperation: (c) common gate pulsing causes similar potentiation (depression) behavior in HT-MoS$_2$ when HB-In$_2$Se$_3$ is potentiated (depressed), (d) multiple P/D cycles of synaptic cooperation between HT-MoS$_2$ and HB-In$_2$Se$_3$. Synaptic competition: (e) modifying $V_{DI}$ from +1 V to -1 V flips the relative strengths of HT-MoS$_2$ and HB-In$_2$Se$_3$ synapses, leading to competitive behavior, (f) multiple P/D cycles of synaptic competition between HT-MoS$_2$ and HB-In$_2$Se$_3$ leading to eventual habituation.}
\label{heterosyn}
\end{figure}

\textbf{Heterosynaptic Plasticity and Synaptic Interactions:} Extending the measurements done on the single synaptic HT-MoS$_2$ FET, the two vertically stacked synapses (HB-In$_2$Se$_3$ and HT-MoS$_2$) of the VSFET can be accessed simultaneously by biasing their S/D terminals, as shown in Figure \ref{heterosyn}a. The source and drain terminals of each channel layer (In$_2$Se$_3$ or MoS$_2$) represent the synaptic ends, and the channel itself represents the synaptic cleft. A single gate terminal simultaneously modulates the conductance of each synaptic channel, beyond the transverse local gate field, wherein, 1) it directly modulates the polarization bound charge and hence the free carrier accumulation and depletion at the hBN/In$_2$Se$_3$ interface of the HB-In$_2$Se$_3$ channel, and 2) it indirectly affects the carrier accumulation at the In$_2$Se$_3$/MoS$_2$ interface through the modulated polarization charge in the In$_2$Se$_3$ layer. This five-terminal measurement of heterosynaptic plasticity, where the plasticity in one input-specific synapse also triggers plasticity in another synapse\cite{chistiakova2014heterosynaptic}, unlike the plain positive feedback homosynaptic plasticity of the HT-MoS$_2$ FET alone, is more useful in maintaining synaptic weight stability, refining synaptic connections and balancing NNs.\cite{chistiakova2009heterosynaptic,jenks2021heterosynaptic}

Simultaneously measured $I_D$ - $V_G$ curves for the HT-MoS$_2$ ($I_{DM}$ - $V_G$, $V_{DM}$ = 1 V) and HB-In$_2$Se$_3$ ($I_{DI}$ - $V_G$, $V_{DI}$ = 1 V) FETs are shown in Figure \ref{heterosyn}b, indicating clockwise hysteresis in both, similar to individually accessed FET characteristics. Further, the VSFET can demonstrate complex but fundamental biological functions such as synaptic cooperation and competition that favor in building and refinement of neural nets.\cite{ramiro2014synaptic} Synaptic cooperation \cite{zhu2019ionic, miller1996synaptic} involves a group of synapses that work together and undergo a similar kind of plasticity change (P/D) in the neural circuit (Figure \ref{heterosyn}c). Applying incremental voltage stimuli at the gate terminal with $V_{DM}$ = 1 V, $V_{DI}$ = 1 V directly modulates the synaptic strength of the bottom HB-In$_2$Se$_3$ memtransistor, which also causes a similar synaptic plasticity response in the upper HT-MoS$_2$ memtransistor. Figure \ref{heterosyn}c shows the simultaneous measured P/D responses of the MoS$_2$ and In$_2$Se$_3$ synapses. Figure \ref{heterosyn}d shows three repeated cycles of P/D data for the two synapses where the MoS$_2$ synapse is stronger than In$_2$Se$_3$, indicating synaptic cooperation. While HT-MoS$_2$ shows a stronger synaptic connection than HB-In$_2$Se$_3$ for all P/D cycles with $V_{DM}$ = 1 V, $V_{DI}$ = 1 V, changing $V_{DI}$ to -1 V, keeping $V_{DM}$ the same, makes the In$_2$Se$_3$ synapse stronger than MoS$_2$ as shown in Figure \ref{heterosyn}e, indicating synaptic competition. Repeating this over three P/D cycles, the HB-In$_2$Se$_3$ synaptic connection weakens and catches up with HT-MoS$_2$ conductance strength, demonstrating a biological habituation response (Figure \ref{heterosyn}f).

%%%%%%%%%%%%figure5

\begin{figure} [H]	
{\includegraphics[width=1\textwidth]{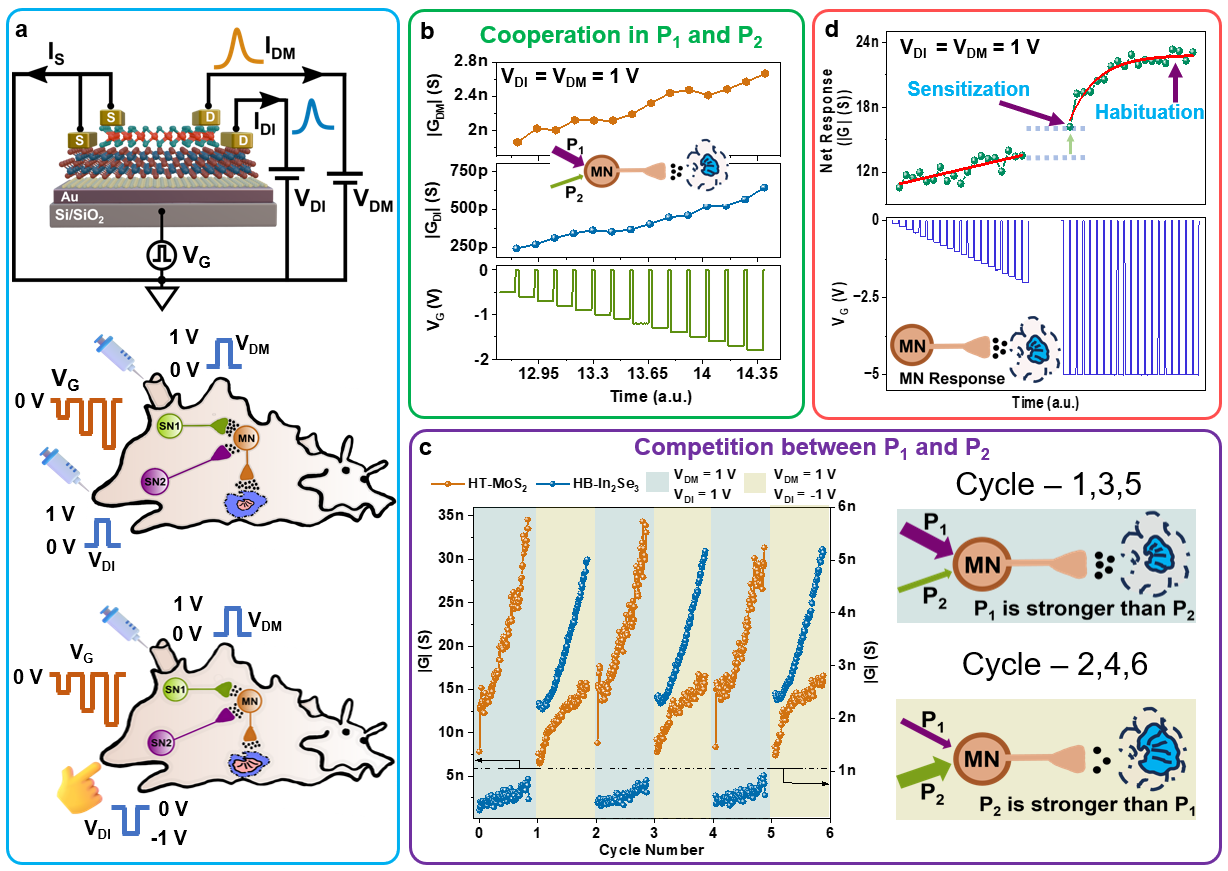}}
\caption{Biomimetic heterosynaptic response and defensive learning behavior of sea mollusk. (a) VSFET electrical circuit equivalence to biological dual-synaptic paths (P$_1$ and P$_2$) of Aplysia sea mollusk connecting its sensory neurons (SN1 and SN2) to the gill motor neuron (MN). The middle and bottom images show the siphon and tail of Aplysia getting stimulated by different tactile (electrical) stimuli. (b) Synaptic cooperative action between P$_1$ and P$_2$ leading to MN-actuated gill withdrawal response. (c) Alternate cycles of synaptic competition between  P$_1$ and P$_2$ resulting in different MN gill responses. (d) Sensitization and habituation forms of learning and training of the gill reflex action to varying stimuli at the gate terminal.}
\label{aplysia}
\end{figure}

 \textbf{\textbf{Biomimetic Heterosynaptic Interactions in Sea Mollusk}:} Heterosynaptic interactions between the HT-MoS$_2$ and HB-MoS$_2$ synapses are shown to mimic the gill withdrawal reflex (GWR) action in Aplysia californica, a marine mollusk. Aplysia’s NN details are given in supporting information 11. Figure \ref{aplysia}a shows the VSFET electrical circuit where the two drain terminals $V_{DM}$ and $V_{DI}$ emulate sensory neurons 1 (SN1) and 2 (SN2), whose PSCs represent the sensory synaptic paths P$_1$ (SN1 $\rightarrow$ MN) and P$_2$ (SN2 $\rightarrow$ MN) to the motor neuron (MN). The gate terminal provides the modulatory stimulus to help the mollusk learn certain behaviors. The middle image of Figure \ref{aplysia}a, where $V_{DM}$ = $V_{D\textit{I}}$ = 1 V, shows a biological analogy to noxious stimuli at the tail and siphon of the mollusk, and the repetitive, pulsed $V_G$ represents a modulatory signal for learning. Likewise, the bottom image in Figure \ref{aplysia}a depicts the change in MN response when the stimulus changes for one sensory neuron ($V_{D\textit{I}}$ = -1 V), for the same input at SN1 and the same modulatory signal. Figure \ref{aplysia}b shows cooperative synaptic interaction between P$_1$ and P$_2$ for harmful stimuli at the two SNs ($V_{DM}$ = $V_{D\textit{I}}$ = 1 V). A small incremental change ($\Delta$$V_G$ = -0.1 V) in the modulatory stimulus potentiates the synaptic strength in both sensory paths leading to MN response manifesting as gill contraction. Here, P$_1$ has stronger synaptic strength compared to P$_2$, and hence, it makes a stronger contribution to MN response than P$_2$. However, when the input to SN2 changes ($V_{DI}$ = -1 V), P$_2$ becomes stronger than P$_1$ (Figure \ref{aplysia}c), depicting competitive behavior. Here, the MN response is slightly weaker. Alternate cycles of cooperation and competition between P$_1$ and P$_2$ are shown in Figure \ref{aplysia}c. Figure \ref{aplysia}d shows how the electrical circuit mimics memory and learning behavior in the mollusk for small incremental $\Delta$$V_G$ = -0.1 V. It first learns slowly how to react to small and incremental inputs ($\Delta$$V_G$ = -0.1 V, $V_G$ = -0.1 V to -2 V) such that when a strong aversive stimulus ($V_G$ = -5 V) comes, it enhances its reflex action due to net synaptic strength enhancement. This refers to sensitization learning behavior in the mollusk, where a surge in neurotransmitter release by the SNs occurs at their synaptic clefts with MNs.\cite{castellucci1976presynaptic} This behavior emerges due to synaptic cooperation. For similar, continued, and redundant strong stimuli at the gate terminal ($V_G$ = -5 V), the motor neuron response slope decreases, showing a decrease in the strength of synaptic connections. This refers to habituation, another form of synaptic competition. Hence, in summary, the gill reflex action of Aplysia mollusk can be efficiently trained through sensitization and habituation forms of learning.\cite{kandel2001molecular}

 %%%%%%%%%%%%figure6
%\newpage

\begin{figure} [H]	
{\includegraphics[width=1\textwidth]{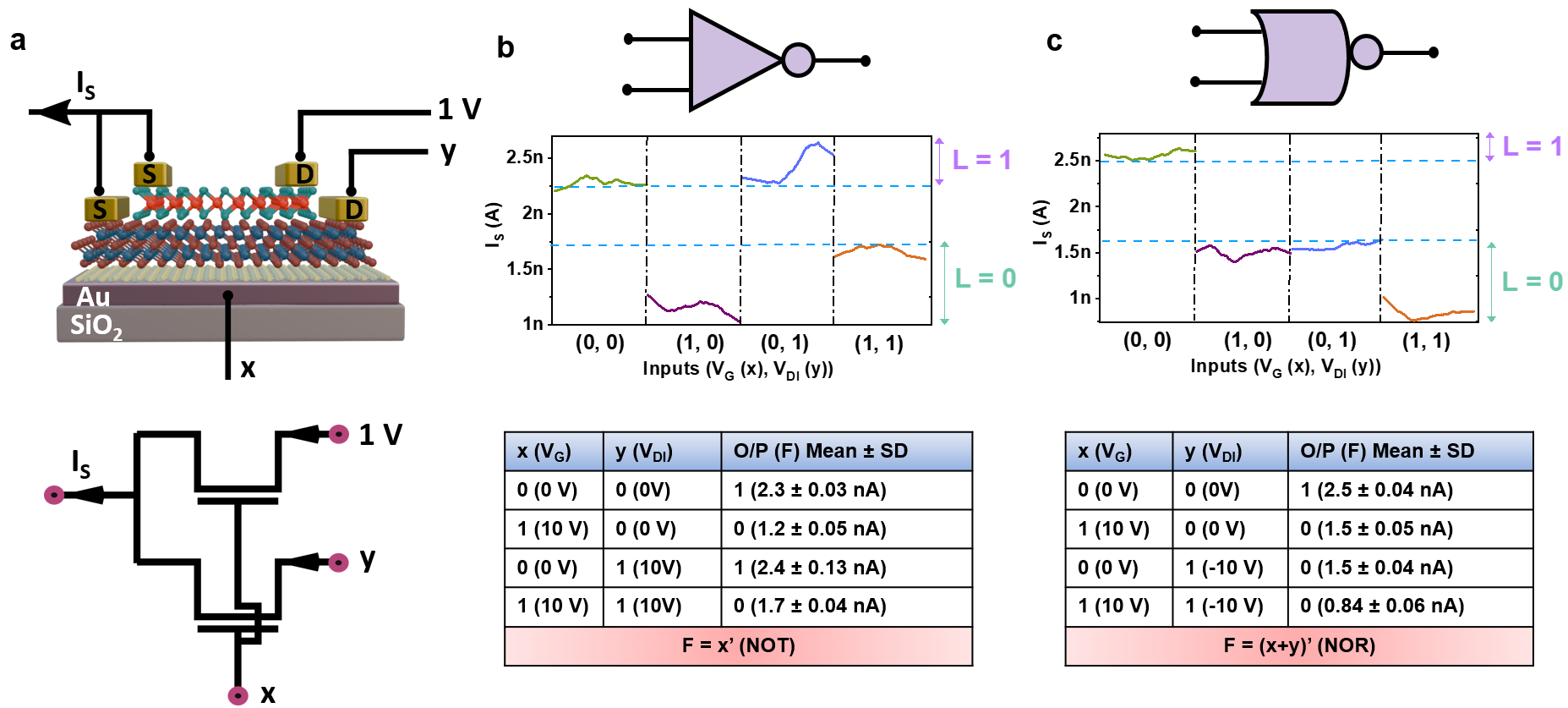}}
\caption{Boolean logic gates using VSFET. (a) 2D device cross-section biasing schematic and equivalent electrical circuit showing the two input variables x and y and the net source output current $I_S$. Output current under different input combinations and the corresponding truth table listing the signal and logic values for (b) NOT, and (c) NOR gate implementation using VSFET.}
\label{logic}
\end{figure}

\textbf{Boolean Logic Gate Functionality:} The VSFET can be easily re-configured to function as a 2-input NOT or NOR logic gate. Figure \ref{logic}a shows the device biasing scheme and equivalent circuit for logic inputs x ($V_G$) and y ($V_{DI}$), where net source current ($I_S$) is the logic output and $V_{DM}$ = 1 V. To realize the NOT gate, as shown in Figure \ref{logic}b, a 0 V at x gives a high output current (output = 1), and 10 V at x gives a low output current (output = 0) irrespective of the voltage applied at y. Likewise, Figure \ref{logic}c shows NOR gate implementation where a 0 V signal applied at both x and y terminals (x = y = 0) results in high $I_S$ (output = 1) while other combinations of 0 V and |10 V| at x and y give a low $I_S$ (output = 0). In summary, VSFET can be reconfigured to realize complete Boolean logic.

%%%%%%%%%%%%figure7

%%%%%%%%%%%%%%%%%%%%%%%%% Conclusion %%%%%%%%%%%%%%%%%%%%%%%%%%%%%%%%%%%%
\section{Conclusion}
This work demonstrates a compact, multifunctional (logic and neuromorphic computing),  and reconfigurable (logic $\rightleftharpoons$ (multi) synaptic memory) VSFET. As shown in Table 1, previous reports of multifunctional and reconfigurable devices are based on trapping, filamentary, phase transition, defects, or electrostatic/ferroelectric gating. The VSFET, on the other hand, leverages the electrostatic coupling between a 2D non-ferroelectric semiconducting MoS$_2$ channel with a 2D ferroelectric semiconducting In$_2$Se$_3$ channel in a vertical heterostructure. Utilizing the out-of-plane In$_2$Se$_3$ ferroelectricity induced non-intrinsic gate-modulated hysteresis in the MoS$_2$ FET, we show homosynaptic plasticity enabled high supervised and unsupervised learning accuracy in artificial and spiking NNs, respectively. Going one step further, simultaneous measurements of gate-controlled memtransistor behaviors of both, the In$_2$Se$_3$ (intrinsic) and the MoS$_2$ (non-intrinsic) channels allow the realization of complex heterosynaptic plasticity behaviors such as synaptic cooperation and competition where the maximum power consumption with both the transistors operational is as low as 6.85 nW at $V_D$ = 0.5 V. These are shown to efficiently mimic the advanced sensorimotor NN controlling gill withdrawal reflex sensitization and habituation in a sea slug (Aplysia). The logic reconfigurability of the VSFET to realize Boolean gates, in addition to its area- and power-efficient multi-memory capabilities, offers significant design flexibility for advanced computing technologies.

%%%%%%%%%%%%%%%%%%%%%%%%%Table1

%\begin{landscape}
\begin {table} [H]
    \centering
    \begin{adjustbox}{width=1\textwidth}
    \Huge
    \newcolumntype{P}[1]{>{\centering\arraybackslash}p{#1}}
    
    \arrayrulecolor{black} % Force border color to black
    \setlength{\arrayrulewidth}{0.7mm}
    \setlength{\tabcolsep}{18pt}
    \renewcommand{\arraystretch}{1.8}
    \begin{tabular}{|p{6.5cm}|P{3.4cm}|P{2.8cm}||P{1.8cm}|P{3cm}|P{2.6cm}|P{3.2cm}|P{2.6cm}|P{3cm}||P{2.4cm}|P{2.8cm}|P{4.6cm}|} 
    \hline
\rowcolor[HTML]{ebf0a4}
\multicolumn{3}{|c||}{} & \multicolumn{6}{c||}{\textbf{Homosynaptic Plasticity}}&\multicolumn{3}{c|}{\textbf{Heterosynaptic Interaction}}\\  \hline 
  
 \cellcolor[HTML]{ffffff}\textbf{Material(s) | Device | Working Principle} & \textbf{Area Efficiency}& \textbf{Recon-figur-ability} &\textbf{P/D}& \textbf{Energy per spike (P/D)} &\textbf{NL} \textbf{$\alpha$$_p$/$\alpha$$_d$ ($\alpha$$_p$ - $\alpha$$_d$)}&\textbf{ANN} \textbf{Accuracy\%}  \textbf{Dataset} 
 \textbf{Epoch}&\textbf{STDP}&\textbf{SNN} \textbf{Accuracy\%}  \textbf{Dataset} 
 \textbf{Epoch}&\textbf{Plast-icity}&\textbf{Coop./\hspace{0cm}Comp.}&\textbf{Biomi-metic}\\ \hline

WSe$_2$ | Split gates homojunction\cite{pan2020reconfigurable} | Electrostatic control&\cellcolor[HTML]{f95c5c}Multi- device Vertical & \cellcolor[HTML]{ace1af}Yes& $ \times $& $ \times $  &$ \times $&$ \times $&\cellcolor[HTML]{ace1af}\checkmark&$ \times $&$ \times $&$ \times $&$ \times $\\ \hline  

In$_2$Se$_3$ | Memresistor\cite{xue2021giant} | Ferroelectric (IP)& \cellcolor[HTML]{f95c5c} Lateral&\cellcolor[HTML]{ace1af}Yes &\cellcolor[HTML]{ace1af}\checkmark&\cellcolor[HTML]{ace1af}8 pJ/\hspace{0cm}8 pJ&$ \times $&\cellcolor[HTML]{ace1af}\checkmark 91 MNIST 30&\cellcolor[HTML]{ace1af}\checkmark&\cellcolor[HTML]{ace1af}\checkmark 89 grey pattern 30&\cellcolor[HTML]{ace1af}\checkmark&\cellcolor[HTML]{ace1af}\checkmark&$ \times $\\ \hline

MoS$_2$ | Memtransistor\cite{sangwan2018multi} | Defect Motion (grain boundaries)&\cellcolor[HTML]{f9bb5c}Vertical&No&\cellcolor[HTML]{ace1af}\checkmark &\cellcolor[HTML]{fdfd96}13.5 nJ/\hspace{0cm}13.5 nJ &$ \times $&$ \times $&\cellcolor[HTML]{ace1af}\checkmark&$ \times $&\cellcolor[HTML]{ace1af}\checkmark&$ \times $&$ \times $\\ \hline  

TaO$_\mathrm{x}$ | Memristor\cite{yang2017multifunctional} | Filamentary&\cellcolor[HTML]{f9bb5c} Vertical&\cellcolor[HTML]{ace1af}Yes&\cellcolor[HTML]{ace1af}\checkmark & \cellcolor[HTML]{ace1af}120 pJ/\hspace{0cm}186 pJ&$ \times $& \cellcolor[HTML]{ace1af}\checkmark&$ \times $&$ \times $ &\cellcolor[HTML]{ace1af}\checkmark&$ \times $&$ \times $\\ \hline

TiO$_{2-\mathrm{x}}$ | Memresistor\cite{miyake2022versatile} | Defect Motion (Vacancies)&\cellcolor[HTML]{f95c5c} Lateral&No&\cellcolor[HTML]{ace1af}\checkmark&\cellcolor[HTML]{f95c5c}6 mJ/6 mJ&$ \times $&$ \times $&\cellcolor[HTML]{ace1af}\checkmark&$ \times $&\cellcolor[HTML]{ace1af}\checkmark&$ \times $&\cellcolor[HTML]{ace1af}\checkmark\\ \hline

Li$_\mathrm{x}$MoO$_3$ | Memresistor\cite{qin2022heterosynaptic} | Ion migration& \cellcolor[HTML]{f95c5c} Lateral&\cellcolor[HTML]{ace1af}Yes&\cellcolor[HTML]{ace1af}\checkmark& $ \times $&$ \times $&$ \times $&$ \times $&$ \times $&\cellcolor[HTML]{ace1af}\checkmark & $ \times $&$ \times $\\ \hline

Li$_\mathrm{x}$MoS$_2$ | Memresistor\cite{zhu2019ionic} | Phase Transition& \cellcolor[HTML]{f95c5c} Lateral&No&\cellcolor[HTML]{ace1af}\checkmark& \cellcolor[HTML]{fdfd96}8 nJ/8 nJ&$ \times $&$ \times $&$ \times $&$ \times $&\cellcolor[HTML]{ace1af}\checkmark&\cellcolor[HTML]{ace1af}\checkmark&$ \times $\\ \hline

WSe$_2$ | Memtransistor\cite{ding2021reconfigurable} | Trapping&\cellcolor[HTML]{f9bb5c} Vertical& No&\cellcolor[HTML]{ace1af}\checkmark&\cellcolor[HTML]{f9bb5c}5.4 $\mu$J/\hspace{0cm}4.725 $\mu$J &\cellcolor[HTML]{ace1af}1/9.6 (-8.6)&$ \times $&$ \times $&$ \times $&\cellcolor[HTML]{ace1af}\checkmark&\cellcolor[HTML]{ace1af}\checkmark&\cellcolor[HTML]{ace1af}\checkmark\\ \hline  

PVP | Memristors\cite{milano2020brain} | Filamentary&\cellcolor[HTML]{f9bb5c} Vertical& No&\cellcolor[HTML]{ace1af}\checkmark&\cellcolor[HTML]{f9825c} 100 $\mu$J/\hspace{0cm}- &$ \times $&$ \times $&$ \times $&$ \times $&\cellcolor[HTML]{ace1af}\checkmark&$ \times $&$ \times $\\ \hline  

Pentacene/APTES | Memtransistor\cite{zheng2020mimicking} | Trapping& \cellcolor[HTML]{fdfd96} Vertical + Lateral&No &\cellcolor[HTML]{ace1af}\checkmark&\cellcolor[HTML]{f9bb5c}1 $\mu$J/\hspace{0cm}675 pJ &$ \times $&\cellcolor[HTML]{ace1af}\checkmark 96.42 MNIST 250&$ \times $&$ \times $&\cellcolor[HTML]{ace1af}\checkmark & \cellcolor[HTML]{ace1af}\checkmark&$ \times $\\ \hline  

MoS$_2$/In$_2$Se$_3$ | Memtransistors | Ferroelectric coupling (OOP) \textbf{This Work}&\cellcolor[HTML]{ace1af}Vertically stratified transistors& \cellcolor[HTML]{ace1af}Yes&\cellcolor[HTML]{ace1af}\checkmark&\cellcolor[HTML]{fdfd96}1.38 nJ/\hspace{0cm}16.8 nJ &\cellcolor[HTML]{ace1af} 1.74/-0.07 (1.81) &\cellcolor[HTML]{ace1af}\checkmark 
90.06 MNIST 5&\cellcolor[HTML]{ace1af}\checkmark&\cellcolor[HTML]{ace1af}\checkmark 96.2 Fisher's Irish 2&\cellcolor[HTML]{ace1af}\checkmark&\cellcolor[HTML]{ace1af}\checkmark&\cellcolor[HTML]{ace1af} \checkmark \\ \hline

   \end{tabular}
    \end{adjustbox}
    \caption{Summary of recently reported area- and energy-efficient reconfigurable devices for multifunctional neuromorphic computing.\cite{xue2021giant,zhu2019ionic,pan2020reconfigurable,sangwan2018multi,yang2017multifunctional,miyake2022versatile,qin2022heterosynaptic,ding2021reconfigurable,milano2020brain,zheng2020mimicking} Here, device reconfigurability indicates easy transitions across the single synapse, multi-synapse and Boolean logic configurations utilizing the same device.}
    \label{benchmarking}
\end{table}
%\end{landscape}

%%%%%%%%%%%%%%%%%%%%%%%%%% Experimental Procedure %%%%%%%%%%%%%%%%%%
\section{Experimental Procedure}

\textbf{Device Fabrication:} Local-gate Au electrodes were first patterned on top of a p$^{+}$ Si/SiO$_2$ (285 nm) substrate with the help of e-beam lithography (EBL, Raith 150-Two) followed by metal (Ti 4 nm/Au 30 nm) deposition (AJA sputter system) and lift-off. Single crystals of MoS$_2$ and In$_2$Se$_3$ were purchased from 2D semiconductors, and hBN crystals were purchased from SPI Supplies. An hBN flake was micromechanically exfoliated using adhesive scotch tape and transferred from the scotch tape to a polydimethylsiloxane (PDMS) stamp. The PDMS stamp was fixed onto a glass slide attached to a micromanipulator. The hBN flake was focus transferred on top of the prefabricated local-gate pattern using the micromanipulator under an Olympus BX-63 microscope. During the transfer process, the Si/SiO$_2$ substrate with the gate pattern was placed on top of a micro heater. After the hBN was aligned and placed on top of the gate pattern, the entire structure, consisting of the Si/SiO$_2$ substrate, the hBN flake along with the PDMS stamp, and the glass slide, was heated up to weaken the adhesion between the PDMS stamp and the hBN flake leaving behind only the hBN flake on top of the gate pattern after PDMS release. The same transfer process was followed for aligning and placing a thick (43.2 nm) In$_2$Se$_3$ flake on top of the hBN flake in the patterned gate area. Next, a rectangular (3.3 $\mu$m  $\times$ 32.6 $\mu$m), thin (4.5 nm) MoS$_2$ flake was transferred on top of the In$_2$Se$_3$ and hBN flakes in such a way that a portion of the MoS$_2$ flake lay on the In$_2$Se$_3$/hBN overlap portion, and the rest on the hBN flake. The overlap portion of MoS$_2$ and In$_2$Se$_3$ flakes forms an MoS$_2$/In$_2$Se$_3$ vertical heterostructure, and the single portions of MoS$_2$ and In$_2$Se$_3$ flakes lying on hBN form the channel layers for control MoS$_2$ and In$_2$Se$_3$ FETs. Next, source/drain contacts were patterned using EBL on the MoS$_2$/In$_2$Se$_3$ heterostructure and the stand-alone  In$_2$Se$_3$ and MoS$_2$ FETs. Finally, contact metallization (Ti 4 nm/Au 100 nm) and lift-off were carried out to complete the device fabrication.

\textbf{Device Characterization:} Electrical measurements were carried out in ambient conditions using a Keysight B1500A semiconductor device analyzer. The steady-state response was measured using Keysight B1500A's HRSMU and MPSMUs, and the pulsed response was measured using Keysight B1500A's WGFMUs.

%%%%%%%%%%%%%%%%%% Supporting Information %%%%%%%%%%%%%%%%%%%%%
%\section{Supporting Information} 
%Supporting Information is available from the Wiley Online Library or from the author.

%%%%%%%%%%%%%%%%%%%%%%%%%%%%%%%%%%%%%%%%%%%%%%%%%%%%%%%%%%%%%%%%%%%%%%%%%%
%% Acknowledgement
%%%%%%%%%%%%%%%%%%%%%%%%%%%%%%%%%%%%%%%%%%%%%%%%%%%%%%%%%%%%%%%%%%%%%
\begin{acknowledgement}
The authors express their gratitude to the Indian Institute of Technology Bombay Nano Fabrication Facility (IITBNF) and the 2D Materials and Devices Lab for providing access to facilities for device fabrication and characterization. S.S. acknowledges the support of the Prime Minister's Research Fellowship (PMRF) PhD scheme, Government of India. S.L. acknowledges funding support from the Department of Science and Technology, under the project grant DST/NM/TUE/QM - 8/2019 (G)/2 and also from project grant FIR/2022/000005 of SERB, Government of India.  
\end{acknowledgement}

%%%%%%%%%%%%%%%%%%%%%%%%%%%%%%%%%%%%%%%%%%%%%%%%%%%%%%%%%%%%%%%%%%%%%%%%%%
%% Conflict of Interest
%%%%%%%%%%%%%%%%%%%%%%%%%%%%%%%%%%%%%%%%%%%%%%%%%%%%%%%%%%%%%%%%%%%%%

%\section{Conflict of Interest}
%The authors declare no conflict of interest.
%%%%%%%%%%%%%%%%%%%%%%%%%%%%%%%%%%%%%%%%%%%%%%%%%%%%%%%%%%%%%%%%%%%%%%%%%%
%% Data Availability Statement
%%%%%%%%%%%%%%%%%%%%%%%%%%%%%%%%%%%%%%%%%%%%%%%%%%%%%%%%%%%%%%%%%%%%%

%\section{Data Availability Statement}
%The data that support the findings of this study are available from the corresponding author upon reasonable request.
%%%%%%%%%%%%%%%%%%%%%%%%%%%%%%%%%%%%%%%%%%%%%%%%%%%%%%%%%%%%%%%%%%%%%
%% Bibliography 
%%%%%%%%%%%%%%%%%%%%%%%%%%%%%%%%%%%%%%%%%%%%%%%%%%%%%%%%%%%%%%%%%%%%%
%\bibliographystyle{MSP}
\bibliography{manuscript}

\providecommand{\latin}[1]{#1}
\makeatletter
\providecommand{\doi}
  {\begingroup\let\do\@makeother\dospecials
  \catcode`\{=1 \catcode`\}=2 \doi@aux}
\providecommand{\doi@aux}[1]{\endgroup\texttt{#1}}
\makeatother
\providecommand*\mcitethebibliography{\thebibliography}
\csname @ifundefined\endcsname{endmcitethebibliography}  {\let\endmcitethebibliography\endthebibliography}{}
\begin{mcitethebibliography}{64}
\providecommand*\natexlab[1]{#1}
\providecommand*\mciteSetBstSublistMode[1]{}
\providecommand*\mciteSetBstMaxWidthForm[2]{}
\providecommand*\mciteBstWouldAddEndPuncttrue
  {\def\EndOfBibitem{\unskip.}}
\providecommand*\mciteBstWouldAddEndPunctfalse
  {\let\EndOfBibitem\relax}
\providecommand*\mciteSetBstMidEndSepPunct[3]{}
\providecommand*\mciteSetBstSublistLabelBeginEnd[3]{}
\providecommand*\EndOfBibitem{}
\mciteSetBstSublistMode{f}
\mciteSetBstMaxWidthForm{subitem}{(\alph{mcitesubitemcount})}
\mciteSetBstSublistLabelBeginEnd
  {\mcitemaxwidthsubitemform\space}
  {\relax}
  {\relax}

\bibitem[Thakar \latin{et~al.}(2023)Thakar, Rajendran, and Lodha]{thakar2023ultra}
Thakar,~K.; Rajendran,~B.; Lodha,~S. Ultra-low power neuromorphic obstacle detection using a two-dimensional materials-based subthreshold transistor. \emph{npj 2D Materials and Applications} \textbf{2023}, \emph{7}, 68\relax
\mciteBstWouldAddEndPuncttrue
\mciteSetBstMidEndSepPunct{\mcitedefaultmidpunct}
{\mcitedefaultendpunct}{\mcitedefaultseppunct}\relax
\EndOfBibitem
\bibitem[Ajayan \latin{et~al.}(2023)Ajayan, Mohankumar, Nirmal, Joseph, Bhattacharya, Sreejith, Kollem, Rebelli, Tayal, and Mounika]{ajayan2023ferroelectric}
Ajayan,~J.; Mohankumar,~P.; Nirmal,~D.; Joseph,~L.~L.; Bhattacharya,~S.; Sreejith,~S.; Kollem,~S.; Rebelli,~S.; Tayal,~S.; Mounika,~B. Ferroelectric field effect transistors (FeFETs): advancements, challenges and exciting prospects for next generation non-volatile memory (NVM) applications. \emph{Materials Today Communications} \textbf{2023}, \emph{35}, 105591\relax
\mciteBstWouldAddEndPuncttrue
\mciteSetBstMidEndSepPunct{\mcitedefaultmidpunct}
{\mcitedefaultendpunct}{\mcitedefaultseppunct}\relax
\EndOfBibitem
\bibitem[Lipatov \latin{et~al.}(2015)Lipatov, Sharma, Gruverman, and Sinitskii]{lipatov2015optoelectrical}
Lipatov,~A.; Sharma,~P.; Gruverman,~A.; Sinitskii,~A. Optoelectrical Molybdenum Disulfide (MoS2)—Ferroelectric Memories. \emph{ACS nano} \textbf{2015}, \emph{9}, 8089--8098\relax
\mciteBstWouldAddEndPuncttrue
\mciteSetBstMidEndSepPunct{\mcitedefaultmidpunct}
{\mcitedefaultendpunct}{\mcitedefaultseppunct}\relax
\EndOfBibitem
\bibitem[Tian \latin{et~al.}(2019)Tian, Liu, Yan, Wang, Zhao, Zhong, Xiang, Sun, Peng, Shen, \latin{et~al.} others]{tian2019robust}
Tian,~B.; Liu,~L.; Yan,~M.; Wang,~J.; Zhao,~Q.; Zhong,~N.; Xiang,~P.; Sun,~L.; Peng,~H.; Shen,~H.; others A robust artificial synapse based on organic ferroelectric polymer. \emph{Advanced Electronic Materials} \textbf{2019}, \emph{5}, 1800600\relax
\mciteBstWouldAddEndPuncttrue
\mciteSetBstMidEndSepPunct{\mcitedefaultmidpunct}
{\mcitedefaultendpunct}{\mcitedefaultseppunct}\relax
\EndOfBibitem
\bibitem[Chen \latin{et~al.}(2020)Chen, Wang, Peng, Feng, Sarkar, Li, Li, Liu, Han, Gong, \latin{et~al.} others]{chen2020van}
Chen,~L.; Wang,~L.; Peng,~Y.; Feng,~X.; Sarkar,~S.; Li,~S.; Li,~B.; Liu,~L.; Han,~K.; Gong,~X.; others A van der Waals synaptic transistor based on ferroelectric Hf0. 5Zr0. 5O2 and 2D tungsten disulfide. \emph{Advanced Electronic Materials} \textbf{2020}, \emph{6}, 2000057\relax
\mciteBstWouldAddEndPuncttrue
\mciteSetBstMidEndSepPunct{\mcitedefaultmidpunct}
{\mcitedefaultendpunct}{\mcitedefaultseppunct}\relax
\EndOfBibitem
\bibitem[Ram \latin{et~al.}(2023)Ram, Maity, Marchand, Mahmoudi, Kshirsagar, Soliman, Taniguchi, Watanabe, Doudin, Ouerghi, \latin{et~al.} others]{ram2023reconfigurable}
Ram,~A.; Maity,~K.; Marchand,~C.; Mahmoudi,~A.; Kshirsagar,~A.~R.; Soliman,~M.; Taniguchi,~T.; Watanabe,~K.; Doudin,~B.; Ouerghi,~A.; others Reconfigurable multifunctional van der Waals ferroelectric devices and logic circuits. \emph{ACS nano} \textbf{2023}, \emph{17}, 21865--21877\relax
\mciteBstWouldAddEndPuncttrue
\mciteSetBstMidEndSepPunct{\mcitedefaultmidpunct}
{\mcitedefaultendpunct}{\mcitedefaultseppunct}\relax
\EndOfBibitem
\bibitem[Wang \latin{et~al.}(2023)Wang, Meng, Wang, Zhang, Li, Yan, Gao, and Ho]{wang20232d}
Wang,~W.; Meng,~Y.; Wang,~W.; Zhang,~Y.; Li,~B.; Yan,~Y.; Gao,~B.; Ho,~J.~C. 2D ferroelectric materials: Emerging paradigms for next-generation ferroelectronics. \emph{Materials Today Electronics} \textbf{2023}, \emph{6}, 100080\relax
\mciteBstWouldAddEndPuncttrue
\mciteSetBstMidEndSepPunct{\mcitedefaultmidpunct}
{\mcitedefaultendpunct}{\mcitedefaultseppunct}\relax
\EndOfBibitem
\bibitem[Baek \latin{et~al.}(2022)Baek, Yoo, Ju, Sriboriboon, Singh, Niu, Park, Shin, Kim, and Lee]{baek2022ferroelectric}
Baek,~S.; Yoo,~H.~H.; Ju,~J.~H.; Sriboriboon,~P.; Singh,~P.; Niu,~J.; Park,~J.-H.; Shin,~C.; Kim,~Y.; Lee,~S. Ferroelectric field-effect-transistor integrated with ferroelectrics heterostructure. \emph{Advanced Science} \textbf{2022}, \emph{9}, 2200566\relax
\mciteBstWouldAddEndPuncttrue
\mciteSetBstMidEndSepPunct{\mcitedefaultmidpunct}
{\mcitedefaultendpunct}{\mcitedefaultseppunct}\relax
\EndOfBibitem
\bibitem[Liu \latin{et~al.}(2016)Liu, You, Seyler, Li, Yu, Lin, Wang, Zhou, Wang, He, \latin{et~al.} others]{liu2016room}
Liu,~F.; You,~L.; Seyler,~K.~L.; Li,~X.; Yu,~P.; Lin,~J.; Wang,~X.; Zhou,~J.; Wang,~H.; He,~H.; others Room-temperature ferroelectricity in CuInP2S6 ultrathin flakes. \emph{Nature communications} \textbf{2016}, \emph{7}, 1--6\relax
\mciteBstWouldAddEndPuncttrue
\mciteSetBstMidEndSepPunct{\mcitedefaultmidpunct}
{\mcitedefaultendpunct}{\mcitedefaultseppunct}\relax
\EndOfBibitem
\bibitem[Wang \latin{et~al.}(2023)Wang, You, Cobden, and Wang]{wang2023towards}
Wang,~C.; You,~L.; Cobden,~D.; Wang,~J. Towards two-dimensional van der Waals ferroelectrics. \emph{Nature Materials} \textbf{2023}, \emph{22}, 542--552\relax
\mciteBstWouldAddEndPuncttrue
\mciteSetBstMidEndSepPunct{\mcitedefaultmidpunct}
{\mcitedefaultendpunct}{\mcitedefaultseppunct}\relax
\EndOfBibitem
\bibitem[Ghosh \latin{et~al.}(2022)Ghosh, Varghese, Jawa, Yin, Medhekar, and Lodha]{ghosh2022polarity}
Ghosh,~S.; Varghese,~A.; Jawa,~H.; Yin,~Y.; Medhekar,~N.~V.; Lodha,~S. Polarity-tunable photocurrent through band alignment engineering in a high-speed WSe2/SnSe2 diode with large negative responsivity. \emph{ACS nano} \textbf{2022}, \emph{16}, 4578--4587\relax
\mciteBstWouldAddEndPuncttrue
\mciteSetBstMidEndSepPunct{\mcitedefaultmidpunct}
{\mcitedefaultendpunct}{\mcitedefaultseppunct}\relax
\EndOfBibitem
\bibitem[Varghese \latin{et~al.}(2020)Varghese, Saha, Thakar, Jindal, Ghosh, Medhekar, Ghosh, and Lodha]{varghese2020near}
Varghese,~A.; Saha,~D.; Thakar,~K.; Jindal,~V.; Ghosh,~S.; Medhekar,~N.~V.; Ghosh,~S.; Lodha,~S. Near-direct bandgap WSe2/ReS2 type-II pn heterojunction for enhanced ultrafast photodetection and high-performance photovoltaics. \emph{Nano letters} \textbf{2020}, \emph{20}, 1707--1717\relax
\mciteBstWouldAddEndPuncttrue
\mciteSetBstMidEndSepPunct{\mcitedefaultmidpunct}
{\mcitedefaultendpunct}{\mcitedefaultseppunct}\relax
\EndOfBibitem
\bibitem[Belianinov \latin{et~al.}(2015)Belianinov, He, Dziaugys, Maksymovych, Eliseev, Borisevich, Morozovska, Banys, Vysochanskii, and Kalinin]{belianinov2015cuinp2s6}
Belianinov,~A.; He,~Q.; Dziaugys,~A.; Maksymovych,~P.; Eliseev,~E.; Borisevich,~A.; Morozovska,~A.; Banys,~J.; Vysochanskii,~Y.; Kalinin,~S.~V. CuInP2S6 room temperature layered ferroelectric. \emph{Nano letters} \textbf{2015}, \emph{15}, 3808--3814\relax
\mciteBstWouldAddEndPuncttrue
\mciteSetBstMidEndSepPunct{\mcitedefaultmidpunct}
{\mcitedefaultendpunct}{\mcitedefaultseppunct}\relax
\EndOfBibitem
\bibitem[Si \latin{et~al.}(2019)Si, Saha, Gao, Qiu, Qin, Duan, Jian, Niu, Wang, Wu, \latin{et~al.} others]{si2019ferroelectric}
Si,~M.; Saha,~A.~K.; Gao,~S.; Qiu,~G.; Qin,~J.; Duan,~Y.; Jian,~J.; Niu,~C.; Wang,~H.; Wu,~W.; others A ferroelectric semiconductor field-effect transistor. \emph{Nature Electronics} \textbf{2019}, \emph{2}, 580--586\relax
\mciteBstWouldAddEndPuncttrue
\mciteSetBstMidEndSepPunct{\mcitedefaultmidpunct}
{\mcitedefaultendpunct}{\mcitedefaultseppunct}\relax
\EndOfBibitem
\bibitem[Liao \latin{et~al.}(2023)Liao, Wen, Wu, Zhou, Hussain, Hu, Li, Liaqat, Zhu, Jiao, \latin{et~al.} others]{liao2023van}
Liao,~J.; Wen,~W.; Wu,~J.; Zhou,~Y.; Hussain,~S.; Hu,~H.; Li,~J.; Liaqat,~A.; Zhu,~H.; Jiao,~L.; others Van der Waals ferroelectric semiconductor field effect transistor for in-memory computing. \emph{ACS nano} \textbf{2023}, \emph{17}, 6095--6102\relax
\mciteBstWouldAddEndPuncttrue
\mciteSetBstMidEndSepPunct{\mcitedefaultmidpunct}
{\mcitedefaultendpunct}{\mcitedefaultseppunct}\relax
\EndOfBibitem
\bibitem[Chang \latin{et~al.}(2020)Chang, Küster, Miller, Ji, Zhang, Sessi, Barraza-Lopez, and Parkin]{chang2020microscopic}
Chang,~K.; Küster,~F.; Miller,~B.~J.; Ji,~J.-R.; Zhang,~J.-L.; Sessi,~P.; Barraza-Lopez,~S.; Parkin,~S.~S. Microscopic manipulation of ferroelectric domains in SnSe monolayers at room temperature. \emph{Nano letters} \textbf{2020}, \emph{20}, 6590--6597\relax
\mciteBstWouldAddEndPuncttrue
\mciteSetBstMidEndSepPunct{\mcitedefaultmidpunct}
{\mcitedefaultendpunct}{\mcitedefaultseppunct}\relax
\EndOfBibitem
\bibitem[Higashitarumizu \latin{et~al.}(2020)Higashitarumizu, Kawamoto, Lee, Lin, Chu, Yonemori, Nishimura, Wakabayashi, Chang, and Nagashio]{higashitarumizu2020purely}
Higashitarumizu,~N.; Kawamoto,~H.; Lee,~C.-J.; Lin,~B.-H.; Chu,~F.-H.; Yonemori,~I.; Nishimura,~T.; Wakabayashi,~K.; Chang,~W.-H.; Nagashio,~K. Purely in-plane ferroelectricity in monolayer SnS at room temperature. \emph{Nature communications} \textbf{2020}, \emph{11}, 2428\relax
\mciteBstWouldAddEndPuncttrue
\mciteSetBstMidEndSepPunct{\mcitedefaultmidpunct}
{\mcitedefaultendpunct}{\mcitedefaultseppunct}\relax
\EndOfBibitem
\bibitem[Bao \latin{et~al.}(2019)Bao, Song, Liu, Chen, Zhu, Abdelwahab, Su, Fu, Chi, Yu, \latin{et~al.} others]{bao2019gate}
Bao,~Y.; Song,~P.; Liu,~Y.; Chen,~Z.; Zhu,~M.; Abdelwahab,~I.; Su,~J.; Fu,~W.; Chi,~X.; Yu,~W.; others Gate-tunable in-plane ferroelectricity in few-layer SnS. \emph{Nano letters} \textbf{2019}, \emph{19}, 5109--5117\relax
\mciteBstWouldAddEndPuncttrue
\mciteSetBstMidEndSepPunct{\mcitedefaultmidpunct}
{\mcitedefaultendpunct}{\mcitedefaultseppunct}\relax
\EndOfBibitem
\bibitem[Xiao \latin{et~al.}(2018)Xiao, Zhu, Wang, Feng, Hu, Dasgupta, Han, Wang, Muller, Martin, \latin{et~al.} others]{xiao2018intrinsic}
Xiao,~J.; Zhu,~H.; Wang,~Y.; Feng,~W.; Hu,~Y.; Dasgupta,~A.; Han,~Y.; Wang,~Y.; Muller,~D.~A.; Martin,~L.~W.; others Intrinsic two-dimensional ferroelectricity with dipole locking. \emph{Physical review letters} \textbf{2018}, \emph{120}, 227601\relax
\mciteBstWouldAddEndPuncttrue
\mciteSetBstMidEndSepPunct{\mcitedefaultmidpunct}
{\mcitedefaultendpunct}{\mcitedefaultseppunct}\relax
\EndOfBibitem
\bibitem[Cui \latin{et~al.}(2018)Cui, Hu, Yan, Addiego, Gao, Wang, Wang, Li, Cheng, Li, \latin{et~al.} others]{cui2018intercorrelated}
Cui,~C.; Hu,~W.-J.; Yan,~X.; Addiego,~C.; Gao,~W.; Wang,~Y.; Wang,~Z.; Li,~L.; Cheng,~Y.; Li,~P.; others Intercorrelated in-plane and out-of-plane ferroelectricity in ultrathin two-dimensional layered semiconductor In2Se3. \emph{Nano letters} \textbf{2018}, \emph{18}, 1253--1258\relax
\mciteBstWouldAddEndPuncttrue
\mciteSetBstMidEndSepPunct{\mcitedefaultmidpunct}
{\mcitedefaultendpunct}{\mcitedefaultseppunct}\relax
\EndOfBibitem
\bibitem[Park \latin{et~al.}(2023)Park, Lee, Kang, Choi, and Park]{park2023laterally}
Park,~S.; Lee,~D.; Kang,~J.; Choi,~H.; Park,~J.-H. Laterally gated ferroelectric field effect transistor (LG-FeFET) using $\alpha$-In2Se3 for stacked in-memory computing array. \emph{Nature Communications} \textbf{2023}, \emph{14}, 6778\relax
\mciteBstWouldAddEndPuncttrue
\mciteSetBstMidEndSepPunct{\mcitedefaultmidpunct}
{\mcitedefaultendpunct}{\mcitedefaultseppunct}\relax
\EndOfBibitem
\bibitem[Wang \latin{et~al.}(2021)Wang, Liu, Gan, Chen, Hou, Ding, Ma, Zhang, and Zhou]{wang2021two}
Wang,~S.; Liu,~L.; Gan,~L.; Chen,~H.; Hou,~X.; Ding,~Y.; Ma,~S.; Zhang,~D.~W.; Zhou,~P. Two-dimensional ferroelectric channel transistors integrating ultra-fast memory and neural computing. \emph{Nature Communications} \textbf{2021}, \emph{12}, 53\relax
\mciteBstWouldAddEndPuncttrue
\mciteSetBstMidEndSepPunct{\mcitedefaultmidpunct}
{\mcitedefaultendpunct}{\mcitedefaultseppunct}\relax
\EndOfBibitem
\bibitem[Kang \latin{et~al.}(2024)Kang, Jung, Gwon, Kim, Byun, Kim, Jang, Shin, Kwon, Cho, \latin{et~al.} others]{kang2024photo}
Kang,~S.-J.; Jung,~W.; Gwon,~O.~H.; Kim,~H.~S.; Byun,~H.~R.; Kim,~J.~Y.; Jang,~S.~G.; Shin,~B.; Kwon,~O.; Cho,~B.; others Photo-Assisted Ferroelectric Domain Control for $\alpha$-In2Se3 Artificial Synapses Inspired by Spontaneous Internal Electric Fields. \emph{Small} \textbf{2024}, \emph{20}, 2307346\relax
\mciteBstWouldAddEndPuncttrue
\mciteSetBstMidEndSepPunct{\mcitedefaultmidpunct}
{\mcitedefaultendpunct}{\mcitedefaultseppunct}\relax
\EndOfBibitem
\bibitem[Kaushik \latin{et~al.}(2014)Kaushik, Nipane, Basheer, Dubey, Grover, Deshmukh, and Lodha]{kaushik2014schottky}
Kaushik,~N.; Nipane,~A.; Basheer,~F.; Dubey,~S.; Grover,~S.; Deshmukh,~M.~M.; Lodha,~S. Schottky barrier heights for Au and Pd contacts to MoS2. \emph{Applied Physics Letters} \textbf{2014}, \emph{105}\relax
\mciteBstWouldAddEndPuncttrue
\mciteSetBstMidEndSepPunct{\mcitedefaultmidpunct}
{\mcitedefaultendpunct}{\mcitedefaultseppunct}\relax
\EndOfBibitem
\bibitem[Jawa \latin{et~al.}(2022)Jawa, Varghese, Ghosh, Sahoo, Yin, Medhekar, and Lodha]{jawa2022wavelength}
Jawa,~H.; Varghese,~A.; Ghosh,~S.; Sahoo,~S.; Yin,~Y.; Medhekar,~N.~V.; Lodha,~S. Wavelength-controlled photocurrent polarity switching in BP-MoS2 heterostructure. \emph{Advanced Functional Materials} \textbf{2022}, \emph{32}, 2112696\relax
\mciteBstWouldAddEndPuncttrue
\mciteSetBstMidEndSepPunct{\mcitedefaultmidpunct}
{\mcitedefaultendpunct}{\mcitedefaultseppunct}\relax
\EndOfBibitem
\bibitem[Yuan \latin{et~al.}(2017)Yuan, Li, Xu, Liu, and Wang]{yuan2017interfacial}
Yuan,~P.; Li,~C.; Xu,~S.; Liu,~J.; Wang,~X. Interfacial thermal conductance between few to tens of layered-MoS2 and c-Si: Effect of MoS2 thickness. \emph{Acta Materialia} \textbf{2017}, \emph{122}, 152--165\relax
\mciteBstWouldAddEndPuncttrue
\mciteSetBstMidEndSepPunct{\mcitedefaultmidpunct}
{\mcitedefaultendpunct}{\mcitedefaultseppunct}\relax
\EndOfBibitem
\bibitem[Lee \latin{et~al.}(2010)Lee, Yan, Brus, Heinz, Hone, and Ryu]{lee2010anomalous}
Lee,~C.; Yan,~H.; Brus,~L.~E.; Heinz,~T.~F.; Hone,~J.; Ryu,~S. Anomalous lattice vibrations of single-and few-layer MoS2. \emph{ACS nano} \textbf{2010}, \emph{4}, 2695--2700\relax
\mciteBstWouldAddEndPuncttrue
\mciteSetBstMidEndSepPunct{\mcitedefaultmidpunct}
{\mcitedefaultendpunct}{\mcitedefaultseppunct}\relax
\EndOfBibitem
\bibitem[Liu \latin{et~al.}(2022)Liu, Zhang, Dang, Bao, Xu, Cheng, Yang, Huang, and Yang]{liu2022optoelectronic}
Liu,~K.; Zhang,~T.; Dang,~B.; Bao,~L.; Xu,~L.; Cheng,~C.; Yang,~Z.; Huang,~R.; Yang,~Y. An optoelectronic synapse based on $\alpha$-In2Se3 with controllable temporal dynamics for multimode and multiscale reservoir computing. \emph{Nature Electronics} \textbf{2022}, \emph{5}, 761--773\relax
\mciteBstWouldAddEndPuncttrue
\mciteSetBstMidEndSepPunct{\mcitedefaultmidpunct}
{\mcitedefaultendpunct}{\mcitedefaultseppunct}\relax
\EndOfBibitem
\bibitem[Dutta \latin{et~al.}(2021)Dutta, Mukherjee, Uzhansky, and Koren]{dutta2021cross}
Dutta,~D.; Mukherjee,~S.; Uzhansky,~M.; Koren,~E. Cross-field optoelectronic modulation via inter-coupled ferroelectricity in 2D In2Se3. \emph{npj 2D Materials and Applications} \textbf{2021}, \emph{5}, 81\relax
\mciteBstWouldAddEndPuncttrue
\mciteSetBstMidEndSepPunct{\mcitedefaultmidpunct}
{\mcitedefaultendpunct}{\mcitedefaultseppunct}\relax
\EndOfBibitem
\bibitem[Xue \latin{et~al.}(2018)Xue, Zhang, Hu, Hsu, Han, Leung, Huang, Wan, Liu, Zhang, \latin{et~al.} others]{xue2018multidirection}
Xue,~F.; Zhang,~J.; Hu,~W.; Hsu,~W.-T.; Han,~A.; Leung,~S.-F.; Huang,~J.-K.; Wan,~Y.; Liu,~S.; Zhang,~J.; others Multidirection piezoelectricity in mono-and multilayered hexagonal $\alpha$-In2Se3. \emph{ACS nano} \textbf{2018}, \emph{12}, 4976--4983\relax
\mciteBstWouldAddEndPuncttrue
\mciteSetBstMidEndSepPunct{\mcitedefaultmidpunct}
{\mcitedefaultendpunct}{\mcitedefaultseppunct}\relax
\EndOfBibitem
\bibitem[Wang \latin{et~al.}(2020)Wang, Wang, Zhang, Li, Ma, Leng, Chen, Abdelwahab, and Loh]{wang2020exploring}
Wang,~L.; Wang,~X.; Zhang,~Y.; Li,~R.; Ma,~T.; Leng,~K.; Chen,~Z.; Abdelwahab,~I.; Loh,~K.~P. Exploring ferroelectric switching in $\alpha$-In2Se3 for neuromorphic computing. \emph{Advanced Functional Materials} \textbf{2020}, \emph{30}, 2004609\relax
\mciteBstWouldAddEndPuncttrue
\mciteSetBstMidEndSepPunct{\mcitedefaultmidpunct}
{\mcitedefaultendpunct}{\mcitedefaultseppunct}\relax
\EndOfBibitem
\bibitem[Varghese \latin{et~al.}(2024)Varghese, Pandey, Sharma, Yin, Medhekar, and Lodha]{varghese2024electrically}
Varghese,~A.; Pandey,~A.~H.; Sharma,~P.; Yin,~Y.; Medhekar,~N.~V.; Lodha,~S. Electrically Controlled High Sensitivity Strain Modulation in MoS2 Field-Effect Transistors via a Piezoelectric Thin Film on Silicon Substrates. \emph{Nano Letters} \textbf{2024}, \emph{24}, 8472--8480\relax
\mciteBstWouldAddEndPuncttrue
\mciteSetBstMidEndSepPunct{\mcitedefaultmidpunct}
{\mcitedefaultendpunct}{\mcitedefaultseppunct}\relax
\EndOfBibitem
\bibitem[Xue \latin{et~al.}(2021)Xue, He, Wang, Retamal, Chai, Jing, Zhang, Fang, Chai, Jiang, \latin{et~al.} others]{xue2021giant}
Xue,~F.; He,~X.; Wang,~Z.; Retamal,~J. R.~D.; Chai,~Z.; Jing,~L.; Zhang,~C.; Fang,~H.; Chai,~Y.; Jiang,~T.; others Giant ferroelectric resistance switching controlled by a modulatory terminal for low-power neuromorphic in-memory computing. \emph{Advanced Materials} \textbf{2021}, \emph{33}, 2008709\relax
\mciteBstWouldAddEndPuncttrue
\mciteSetBstMidEndSepPunct{\mcitedefaultmidpunct}
{\mcitedefaultendpunct}{\mcitedefaultseppunct}\relax
\EndOfBibitem
\bibitem[Jawa \latin{et~al.}(2021)Jawa, Varghese, and Lodha]{jawa2021electrically}
Jawa,~H.; Varghese,~A.; Lodha,~S. Electrically tunable room temperature hysteresis crossover in underlap MoS2 field-effect transistors. \emph{ACS Applied Materials \& Interfaces} \textbf{2021}, \emph{13}, 9186--9194\relax
\mciteBstWouldAddEndPuncttrue
\mciteSetBstMidEndSepPunct{\mcitedefaultmidpunct}
{\mcitedefaultendpunct}{\mcitedefaultseppunct}\relax
\EndOfBibitem
\bibitem[Mu \latin{et~al.}(2024)Mu, Yang, Xie, Wang, Guo, and Gong]{mu2024homo}
Mu,~Y.; Yang,~J.; Xie,~G.; Wang,~Z.; Guo,~B.; Gong,~J.~R. Homo-type $\alpha$-In2Se3/PdSe2 Ferroelectric van der Waals Heterojunction Photodetectors with High-performance and Broadband. \emph{Advanced Functional Materials} \textbf{2024}, 2315543\relax
\mciteBstWouldAddEndPuncttrue
\mciteSetBstMidEndSepPunct{\mcitedefaultmidpunct}
{\mcitedefaultendpunct}{\mcitedefaultseppunct}\relax
\EndOfBibitem
\bibitem[Lyu \latin{et~al.}(2020)Lyu, Sun, Yang, Tang, Li, Li, Sun, Gao, Ye, and Chen]{lyu2020thickness}
Lyu,~F.; Sun,~Y.; Yang,~Q.; Tang,~B.; Li,~M.; Li,~Z.; Sun,~M.; Gao,~P.; Ye,~L.-H.; Chen,~Q. Thickness-dependent band gap of $\alpha$-In2Se3: from electron energy loss spectroscopy to density functional theory calculations. \emph{Nanotechnology} \textbf{2020}, \emph{31}, 315711\relax
\mciteBstWouldAddEndPuncttrue
\mciteSetBstMidEndSepPunct{\mcitedefaultmidpunct}
{\mcitedefaultendpunct}{\mcitedefaultseppunct}\relax
\EndOfBibitem
\bibitem[Elias \latin{et~al.}(2019)Elias, Valvin, Pelini, Summerfield, Mellor, Cheng, Eaves, Foxon, Beton, Novikov, \latin{et~al.} others]{elias2019direct}
Elias,~C.; Valvin,~P.; Pelini,~T.; Summerfield,~A.; Mellor,~C.; Cheng,~T.; Eaves,~L.; Foxon,~C.; Beton,~P.; Novikov,~S.; others Direct band-gap crossover in epitaxial monolayer boron nitride. \emph{Nature communications} \textbf{2019}, \emph{10}, 2639\relax
\mciteBstWouldAddEndPuncttrue
\mciteSetBstMidEndSepPunct{\mcitedefaultmidpunct}
{\mcitedefaultendpunct}{\mcitedefaultseppunct}\relax
\EndOfBibitem
\bibitem[Cassabois \latin{et~al.}(2016)Cassabois, Valvin, and Gil]{cassabois2016hexagonal}
Cassabois,~G.; Valvin,~P.; Gil,~B. Hexagonal boron nitride is an indirect bandgap semiconductor. \emph{Nature photonics} \textbf{2016}, \emph{10}, 262--266\relax
\mciteBstWouldAddEndPuncttrue
\mciteSetBstMidEndSepPunct{\mcitedefaultmidpunct}
{\mcitedefaultendpunct}{\mcitedefaultseppunct}\relax
\EndOfBibitem
\bibitem[Kaushik \latin{et~al.}(2020)Kaushik, Sharda, and Bhowmik]{kaushik2020synapse}
Kaushik,~D.; Sharda,~J.; Bhowmik,~D. Synapse cell optimization and back-propagation algorithm implementation in a domain wall synapse based crossbar neural network for scalable on-chip learning. \emph{Nanotechnology} \textbf{2020}, \emph{31}, 364004\relax
\mciteBstWouldAddEndPuncttrue
\mciteSetBstMidEndSepPunct{\mcitedefaultmidpunct}
{\mcitedefaultendpunct}{\mcitedefaultseppunct}\relax
\EndOfBibitem
\bibitem[Yadav \latin{et~al.}(2023)Yadav, Gupta, Holla, Ali~Khan, Muduli, and Bhowmik]{yadav2023demonstration}
Yadav,~R.~S.; Gupta,~P.; Holla,~A.; Ali~Khan,~K.~I.; Muduli,~P.~K.; Bhowmik,~D. Demonstration of Synaptic Behavior in a Heavy-Metal-Ferromagnetic-Metal-Oxide-Heterostructure-Based Spintronic Device for On-Chip Learning in Crossbar-Array-Based Neural Networks. \emph{ACS Applied Electronic Materials} \textbf{2023}, \emph{5}, 484--497\relax
\mciteBstWouldAddEndPuncttrue
\mciteSetBstMidEndSepPunct{\mcitedefaultmidpunct}
{\mcitedefaultendpunct}{\mcitedefaultseppunct}\relax
\EndOfBibitem
\bibitem[Hwang \latin{et~al.}(2023)Hwang, Park, Hwang, Choi, Kim, Kim, Choi, Yoon, Kwon, and Kim]{hwang2023bioinspired}
Hwang,~Y.; Park,~B.; Hwang,~S.; Choi,~S.-W.; Kim,~H.~S.; Kim,~A.~R.; Choi,~J.~W.; Yoon,~J.; Kwon,~J.-D.; Kim,~Y. A Bioinspired Ultra Flexible Artificial van der Waals 2D-MoS2 Channel/LiSiOx Solid Electrolyte Synapse Arrays via Laser-Lift Off Process for Wearable Adaptive Neuromorphic Computing. \emph{Small Methods} \textbf{2023}, \emph{7}, 2201719\relax
\mciteBstWouldAddEndPuncttrue
\mciteSetBstMidEndSepPunct{\mcitedefaultmidpunct}
{\mcitedefaultendpunct}{\mcitedefaultseppunct}\relax
\EndOfBibitem
\bibitem[Crair and Shah(2009)Crair, and Shah]{crair2009long}
Crair,~M.; Shah,~R. Long-term potentiation and long-term depression in experience-dependent plasticity. \textbf{2009}, \relax
\mciteBstWouldAddEndPunctfalse
\mciteSetBstMidEndSepPunct{\mcitedefaultmidpunct}
{}{\mcitedefaultseppunct}\relax
\EndOfBibitem
\bibitem[G{\"u}tig(2014)]{gutig2014spike}
G{\"u}tig,~R. To spike, or when to spike? \emph{Current opinion in neurobiology} \textbf{2014}, \emph{25}, 134--139\relax
\mciteBstWouldAddEndPuncttrue
\mciteSetBstMidEndSepPunct{\mcitedefaultmidpunct}
{\mcitedefaultendpunct}{\mcitedefaultseppunct}\relax
\EndOfBibitem
\bibitem[Sahu \latin{et~al.}(2019)Sahu, Pandey, Goyal, and Bhowmik]{sahu2019spike}
Sahu,~U.; Pandey,~A.; Goyal,~K.; Bhowmik,~D. Spike time dependent plasticity (STDP) enabled learning in spiking neural networks using domain wall based synapses and neurons. \emph{AIP Advances} \textbf{2019}, \emph{9}\relax
\mciteBstWouldAddEndPuncttrue
\mciteSetBstMidEndSepPunct{\mcitedefaultmidpunct}
{\mcitedefaultendpunct}{\mcitedefaultseppunct}\relax
\EndOfBibitem
\bibitem[Stang(2010)]{iris_setosa_image}
Stang,~D.~J. Iris setosa 2. \url{https://commons.wikimedia.org/wiki/File:Iris_setosa_2.jpg}, 2010; Accessed: 2024-07-01\relax
\mciteBstWouldAddEndPuncttrue
\mciteSetBstMidEndSepPunct{\mcitedefaultmidpunct}
{\mcitedefaultendpunct}{\mcitedefaultseppunct}\relax
\EndOfBibitem
\bibitem[Mayfield(2007)]{iris_versicolor_image}
Mayfield,~F. Iris versicolor 3. \url{https://commons.wikimedia.org/wiki/File:Iris_versicolor_3.jpg}, 2007; Accessed: 2024-07-01\relax
\mciteBstWouldAddEndPuncttrue
\mciteSetBstMidEndSepPunct{\mcitedefaultmidpunct}
{\mcitedefaultendpunct}{\mcitedefaultseppunct}\relax
\EndOfBibitem
\bibitem[Richards(2009)]{iris_virginica_image}
Richards,~F.~D. Iris virginica. \url{https://commons.wikimedia.org/wiki/File:Iris_virginica.jpg}, 2009; Accessed: 2024-07-01\relax
\mciteBstWouldAddEndPuncttrue
\mciteSetBstMidEndSepPunct{\mcitedefaultmidpunct}
{\mcitedefaultendpunct}{\mcitedefaultseppunct}\relax
\EndOfBibitem
\bibitem[Chistiakova \latin{et~al.}(2014)Chistiakova, Bannon, Bazhenov, and Volgushev]{chistiakova2014heterosynaptic}
Chistiakova,~M.; Bannon,~N.~M.; Bazhenov,~M.; Volgushev,~M. Heterosynaptic plasticity: multiple mechanisms and multiple roles. \emph{The Neuroscientist} \textbf{2014}, \emph{20}, 483--498\relax
\mciteBstWouldAddEndPuncttrue
\mciteSetBstMidEndSepPunct{\mcitedefaultmidpunct}
{\mcitedefaultendpunct}{\mcitedefaultseppunct}\relax
\EndOfBibitem
\bibitem[Chistiakova and Volgushev(2009)Chistiakova, and Volgushev]{chistiakova2009heterosynaptic}
Chistiakova,~M.; Volgushev,~M. Heterosynaptic plasticity in the neocortex. \emph{Experimental brain research} \textbf{2009}, \emph{199}, 377--390\relax
\mciteBstWouldAddEndPuncttrue
\mciteSetBstMidEndSepPunct{\mcitedefaultmidpunct}
{\mcitedefaultendpunct}{\mcitedefaultseppunct}\relax
\EndOfBibitem
\bibitem[Jenks \latin{et~al.}(2021)Jenks, Tsimring, Ip, Zepeda, and Sur]{jenks2021heterosynaptic}
Jenks,~K.~R.; Tsimring,~K.; Ip,~J. P.~K.; Zepeda,~J.~C.; Sur,~M. Heterosynaptic plasticity and the experience-dependent refinement of developing neuronal circuits. \emph{Frontiers in neural circuits} \textbf{2021}, \emph{15}, 803401\relax
\mciteBstWouldAddEndPuncttrue
\mciteSetBstMidEndSepPunct{\mcitedefaultmidpunct}
{\mcitedefaultendpunct}{\mcitedefaultseppunct}\relax
\EndOfBibitem
\bibitem[Ramiro-Cort{\'e}s \latin{et~al.}(2014)Ramiro-Cort{\'e}s, Hobbiss, and Israely]{ramiro2014synaptic}
Ramiro-Cort{\'e}s,~Y.; Hobbiss,~A.~F.; Israely,~I. Synaptic competition in structural plasticity and cognitive function. \emph{Philosophical Transactions of the Royal Society B: Biological Sciences} \textbf{2014}, \emph{369}, 20130157\relax
\mciteBstWouldAddEndPuncttrue
\mciteSetBstMidEndSepPunct{\mcitedefaultmidpunct}
{\mcitedefaultendpunct}{\mcitedefaultseppunct}\relax
\EndOfBibitem
\bibitem[Zhu \latin{et~al.}(2019)Zhu, Li, Liang, and Lu]{zhu2019ionic}
Zhu,~X.; Li,~D.; Liang,~X.; Lu,~W.~D. Ionic modulation and ionic coupling effects in MoS2 devices for neuromorphic computing. \emph{Nature materials} \textbf{2019}, \emph{18}, 141--148\relax
\mciteBstWouldAddEndPuncttrue
\mciteSetBstMidEndSepPunct{\mcitedefaultmidpunct}
{\mcitedefaultendpunct}{\mcitedefaultseppunct}\relax
\EndOfBibitem
\bibitem[Miller(1996)]{miller1996synaptic}
Miller,~K.~D. Synaptic economics: competition and cooperation in synaptic plasticity. \emph{Neuron} \textbf{1996}, \emph{17}, 371--374\relax
\mciteBstWouldAddEndPuncttrue
\mciteSetBstMidEndSepPunct{\mcitedefaultmidpunct}
{\mcitedefaultendpunct}{\mcitedefaultseppunct}\relax
\EndOfBibitem
\bibitem[Castellucci and Kandel(1976)Castellucci, and Kandel]{castellucci1976presynaptic}
Castellucci,~V.; Kandel,~E.~R. Presynaptic facilitation as a mechanism for behavioral sensitization in Aplysia. \emph{Science} \textbf{1976}, \emph{194}, 1176--1178\relax
\mciteBstWouldAddEndPuncttrue
\mciteSetBstMidEndSepPunct{\mcitedefaultmidpunct}
{\mcitedefaultendpunct}{\mcitedefaultseppunct}\relax
\EndOfBibitem
\bibitem[Kandel(2001)]{kandel2001molecular}
Kandel,~E.~R. The molecular biology of memory storage: a dialogue between genes and synapses. \emph{Science} \textbf{2001}, \emph{294}, 1030--1038\relax
\mciteBstWouldAddEndPuncttrue
\mciteSetBstMidEndSepPunct{\mcitedefaultmidpunct}
{\mcitedefaultendpunct}{\mcitedefaultseppunct}\relax
\EndOfBibitem
\bibitem[Pan \latin{et~al.}(2020)Pan, Wang, Liang, Wang, Cao, Wang, Wang, Wang, Cheng, Gao, \latin{et~al.} others]{pan2020reconfigurable}
Pan,~C.; Wang,~C.-Y.; Liang,~S.-J.; Wang,~Y.; Cao,~T.; Wang,~P.; Wang,~C.; Wang,~S.; Cheng,~B.; Gao,~A.; others Reconfigurable logic and neuromorphic circuits based on electrically tunable two-dimensional homojunctions. \emph{Nature Electronics} \textbf{2020}, \emph{3}, 383--390\relax
\mciteBstWouldAddEndPuncttrue
\mciteSetBstMidEndSepPunct{\mcitedefaultmidpunct}
{\mcitedefaultendpunct}{\mcitedefaultseppunct}\relax
\EndOfBibitem
\bibitem[Sangwan \latin{et~al.}(2018)Sangwan, Lee, Bergeron, Balla, Beck, Chen, and Hersam]{sangwan2018multi}
Sangwan,~V.~K.; Lee,~H.-S.; Bergeron,~H.; Balla,~I.; Beck,~M.~E.; Chen,~K.-S.; Hersam,~M.~C. Multi-terminal memtransistors from polycrystalline monolayer molybdenum disulfide. \emph{Nature} \textbf{2018}, \emph{554}, 500--504\relax
\mciteBstWouldAddEndPuncttrue
\mciteSetBstMidEndSepPunct{\mcitedefaultmidpunct}
{\mcitedefaultendpunct}{\mcitedefaultseppunct}\relax
\EndOfBibitem
\bibitem[Yang \latin{et~al.}(2017)Yang, Yin, Yu, Wang, Zhang, Cai, Lu, and Huang]{yang2017multifunctional}
Yang,~Y.; Yin,~M.; Yu,~Z.; Wang,~Z.; Zhang,~T.; Cai,~Y.; Lu,~W.~D.; Huang,~R. Multifunctional Nanoionic Devices Enabling Simultaneous Heterosynaptic Plasticity and Efficient In-Memory Boolean Logic. \emph{Advanced Electronic Materials} \textbf{2017}, \emph{3}, 1700032\relax
\mciteBstWouldAddEndPuncttrue
\mciteSetBstMidEndSepPunct{\mcitedefaultmidpunct}
{\mcitedefaultendpunct}{\mcitedefaultseppunct}\relax
\EndOfBibitem
\bibitem[Miyake \latin{et~al.}(2022)Miyake, Nagata, Adachi, Hayashi, Tohei, and Sakai]{miyake2022versatile}
Miyake,~R.; Nagata,~Z.; Adachi,~K.; Hayashi,~Y.; Tohei,~T.; Sakai,~A. Versatile functionality of four-terminal TiO2--x memristive devices as artificial synapses for neuromorphic computing. \emph{ACS Applied Electronic Materials} \textbf{2022}, \emph{4}, 2326--2336\relax
\mciteBstWouldAddEndPuncttrue
\mciteSetBstMidEndSepPunct{\mcitedefaultmidpunct}
{\mcitedefaultendpunct}{\mcitedefaultseppunct}\relax
\EndOfBibitem
\bibitem[Qin \latin{et~al.}(2022)Qin, Zhu, Wang, Zhu, Sun, Zhen, Chai, and Xu]{qin2022heterosynaptic}
Qin,~J.-K.; Zhu,~B.-X.; Wang,~C.; Zhu,~C.-Y.; Sun,~R.-Y.; Zhen,~L.; Chai,~Y.; Xu,~C.-Y. Heterosynaptic plasticity achieved by highly anisotropic ionic migration in layered LixMoO3 for neuromorphic application. \emph{Advanced Electronic Materials} \textbf{2022}, \emph{8}, 2200721\relax
\mciteBstWouldAddEndPuncttrue
\mciteSetBstMidEndSepPunct{\mcitedefaultmidpunct}
{\mcitedefaultendpunct}{\mcitedefaultseppunct}\relax
\EndOfBibitem
\bibitem[Ding \latin{et~al.}(2021)Ding, Yang, Chen, Mo, Zhou, Liu, Shang, Zhai, Han, and Zhou]{ding2021reconfigurable}
Ding,~G.; Yang,~B.; Chen,~R.-S.; Mo,~W.-A.; Zhou,~K.; Liu,~Y.; Shang,~G.; Zhai,~Y.; Han,~S.-T.; Zhou,~Y. Reconfigurable 2D WSe2-based memtransistor for mimicking homosynaptic and heterosynaptic plasticity. \emph{Small} \textbf{2021}, \emph{17}, 2103175\relax
\mciteBstWouldAddEndPuncttrue
\mciteSetBstMidEndSepPunct{\mcitedefaultmidpunct}
{\mcitedefaultendpunct}{\mcitedefaultseppunct}\relax
\EndOfBibitem
\bibitem[Milano \latin{et~al.}(2020)Milano, Pedretti, Fretto, Boarino, Benfenati, Ielmini, Valov, and Ricciardi]{milano2020brain}
Milano,~G.; Pedretti,~G.; Fretto,~M.; Boarino,~L.; Benfenati,~F.; Ielmini,~D.; Valov,~I.; Ricciardi,~C. Brain-inspired structural plasticity through reweighting and rewiring in multi-terminal self-organizing memristive nanowire networks. \emph{Advanced Intelligent Systems} \textbf{2020}, \emph{2}, 2000096\relax
\mciteBstWouldAddEndPuncttrue
\mciteSetBstMidEndSepPunct{\mcitedefaultmidpunct}
{\mcitedefaultendpunct}{\mcitedefaultseppunct}\relax
\EndOfBibitem
\bibitem[Zheng \latin{et~al.}(2020)Zheng, Liao, Xiong, Zhou, and Han]{zheng2020mimicking}
Zheng,~C.; Liao,~Y.; Xiong,~Z.; Zhou,~Y.; Han,~S.-T. Mimicking the competitive and cooperative behaviors with multi-terminal synaptic memtransistors. \emph{Journal of Materials Chemistry C} \textbf{2020}, \emph{8}, 6063--6071\relax
\mciteBstWouldAddEndPuncttrue
\mciteSetBstMidEndSepPunct{\mcitedefaultmidpunct}
{\mcitedefaultendpunct}{\mcitedefaultseppunct}\relax
\EndOfBibitem
\end{mcitethebibliography}


\providecommand{\latin}[1]{#1}
\makeatletter
\providecommand{\doi}
  {\begingroup\let\do\@makeother\dospecials
  \catcode`\{=1 \catcode`\}=2 \doi@aux}
\providecommand{\doi@aux}[1]{\endgroup\texttt{#1}}
\makeatother
\providecommand*\mcitethebibliography{\thebibliography}
\csname @ifundefined\endcsname{endmcitethebibliography}  {\let\endmcitethebibliography\endthebibliography}{}
\begin{mcitethebibliography}{5}
\providecommand*\natexlab[1]{#1}
\providecommand*\mciteSetBstSublistMode[1]{}
\providecommand*\mciteSetBstMaxWidthForm[2]{}
\providecommand*\mciteBstWouldAddEndPuncttrue
  {\def\EndOfBibitem{\unskip.}}
\providecommand*\mciteBstWouldAddEndPunctfalse
  {\let\EndOfBibitem\relax}
\providecommand*\mciteSetBstMidEndSepPunct[3]{}
\providecommand*\mciteSetBstSublistLabelBeginEnd[3]{}
\providecommand*\EndOfBibitem{}
\mciteSetBstSublistMode{f}
\mciteSetBstMaxWidthForm{subitem}{(\alph{mcitesubitemcount})}
\mciteSetBstSublistLabelBeginEnd
  {\mcitemaxwidthsubitemform\space}
  {\relax}
  {\relax}

\bibitem[Contributors(2024)]{NeuroSim2024}
Contributors,~N. DNN NeuroSim V2.0 Nonlinearity Norms Documentation. \url{https://github.com/neurosim/DNN_NeuroSim_V2.0/blob/master/Documents/Nonlinearity-NormA.htm}, 2024; Accessed: 26 July 2024\relax
\mciteBstWouldAddEndPuncttrue
\mciteSetBstMidEndSepPunct{\mcitedefaultmidpunct}
{\mcitedefaultendpunct}{\mcitedefaultseppunct}\relax
\EndOfBibitem
\bibitem[Moroz(2011)]{moroz2011aplysia}
Moroz,~L.~L. Aplysia. \emph{Current Biology} \textbf{2011}, \emph{21}, R60--R61\relax
\mciteBstWouldAddEndPuncttrue
\mciteSetBstMidEndSepPunct{\mcitedefaultmidpunct}
{\mcitedefaultendpunct}{\mcitedefaultseppunct}\relax
\EndOfBibitem
\bibitem[Pinsker \latin{et~al.}(1970)Pinsker, Kupfermann, Castellucci, and Kandel]{pinsker1970habituation}
Pinsker,~H.; Kupfermann,~I.; Castellucci,~V.; Kandel,~E. Habituation and dishabituation of the gill-withdrawal reflex in Aplysia. \emph{Science} \textbf{1970}, \emph{167}, 1740--1742\relax
\mciteBstWouldAddEndPuncttrue
\mciteSetBstMidEndSepPunct{\mcitedefaultmidpunct}
{\mcitedefaultendpunct}{\mcitedefaultseppunct}\relax
\EndOfBibitem
\bibitem[Chitwood \latin{et~al.}(2001)Chitwood, Li, and Glanzman]{chitwood2001serotonin}
Chitwood,~R.~A.; Li,~Q.; Glanzman,~D.~L. Serotonin facilitates AMPA-type responses in isolated siphon motor neurons of Aplysia in culture. \emph{The Journal of physiology} \textbf{2001}, \emph{534}, 501\relax
\mciteBstWouldAddEndPuncttrue
\mciteSetBstMidEndSepPunct{\mcitedefaultmidpunct}
{\mcitedefaultendpunct}{\mcitedefaultseppunct}\relax
\EndOfBibitem
\end{mcitethebibliography}
\end{document}

\setcitestyle{square} % Use square brackets in in-text citations

\makeatletter
% Use square brackets in bibliography
\renewcommand*{\@biblabel}[1]{[#1]}
\makeatother

\renewcommand{\thefigure}{S\arabic{figure}}
\renewcommand{\theequation}{S\arabic{equation}}

%\begin{My_Supporting_Info}

%%%%%%%%%%%%Supporting figure1
\newpage
\textbf{Supporting Information 1:} \textbf{2D cross-sections of C-MoS$_2$ and C-In$_2$Se$_3$ FETs and PFM characterization.}
\begin{figure}[H]	
{\includegraphics[width=1\textwidth]{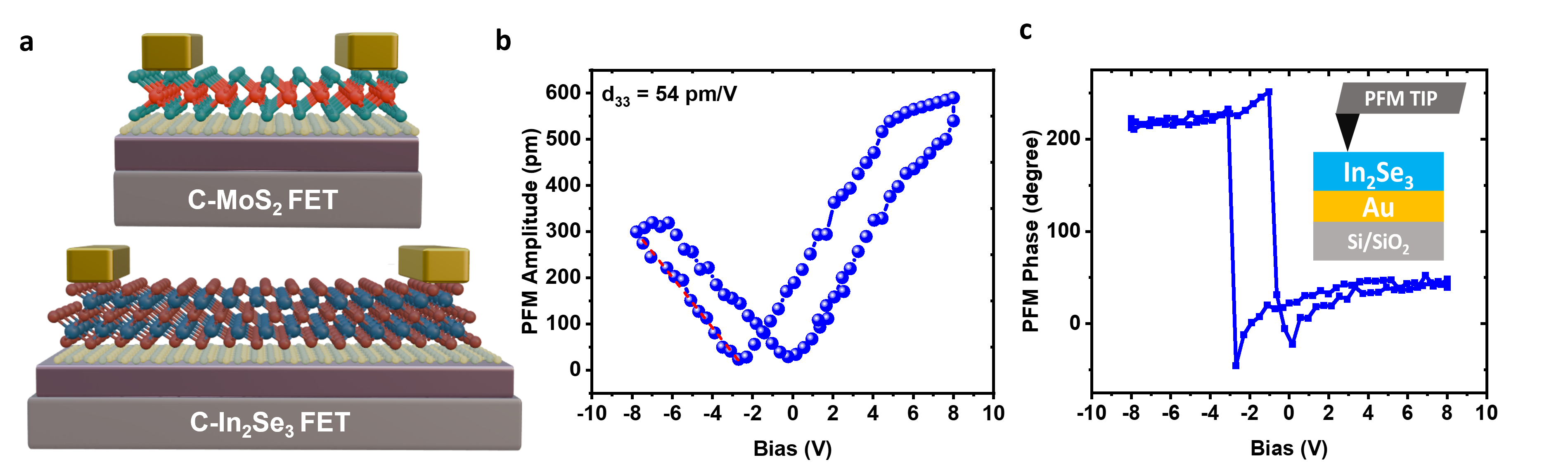}}
	\quad
\caption{(a) 2D cross-sections of control MoS$_2$ FET (C-MoS$_2$ FET) and control In$_2$Se$_3$ FET (C-In$_2$Se$_3$ FET). Piezoresponse force microscopy (PFM) characterization of In$_2$Se$_3$ flake showing (b) PFM amplitude, and (c) PFM phase versus voltage hysteresis loop indicating clear ferroelectric polarization switching under an external electric field. The inset shows the schematic of the PFM measurement.}
\label{s1}
\end{figure}

%%%%%%%%%%%%Supporting figure2
\newpage
\textbf{Supporting Information 2:} \textbf{Double sweep transfer ($I_D$ - $V_G$) characteristics at $V_D$ = 1 V and extracted memory window (MW)/sweep range (SR) ratios of two additional HT-MoS$_2$ FETs at $V_D$ = 1 V.}
\begin{figure}[H]	
{\includegraphics[width=1\textwidth]{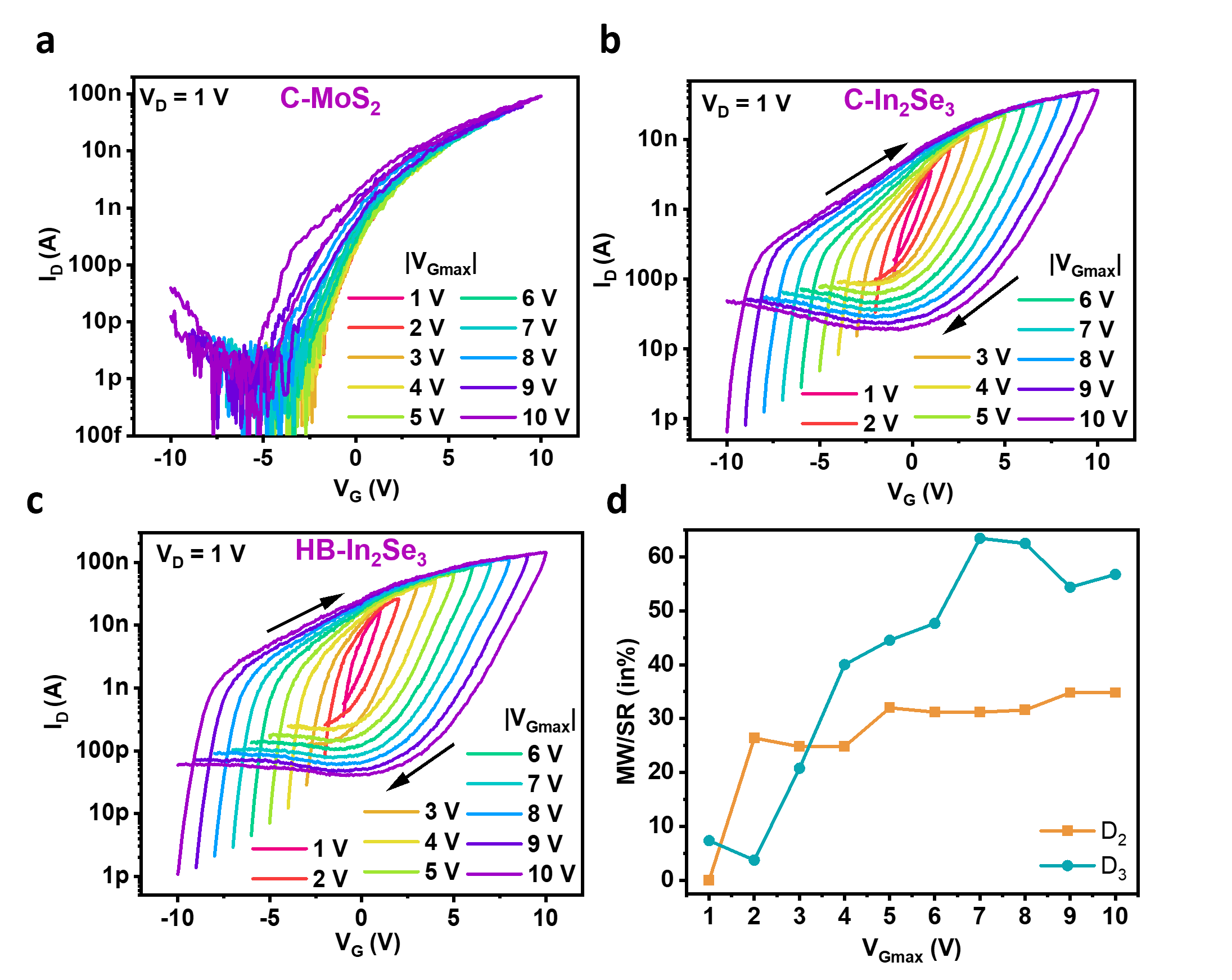}}
	\quad
\caption{ Double sweep transfer ($I_D$ - $V_G$) characteristics of (a) C-MoS$_2$ FET, (b) C-In$_2$Se$_3$ FET, and (b) HB-In$_2$Se$_3$ FET for different maximum gate bias values. (d) MW/SR ratios extracted as a function of gate bias sweep range of two additional HT-MoS$_2$ FETs at $V_D$ = 1 V. }
\label{s2}
\end{figure}

%%%%%%%%%%%%Supporting figure
\newpage
\textbf{Supporting Information 3:} \textbf{Double sweep transfer ($I_D$ - $V_G$) characteristics  and the extracted MW/SR ratios at $V_D$ = 0.5 V.}
\begin{figure}[H]	
{\includegraphics[width=1\textwidth]{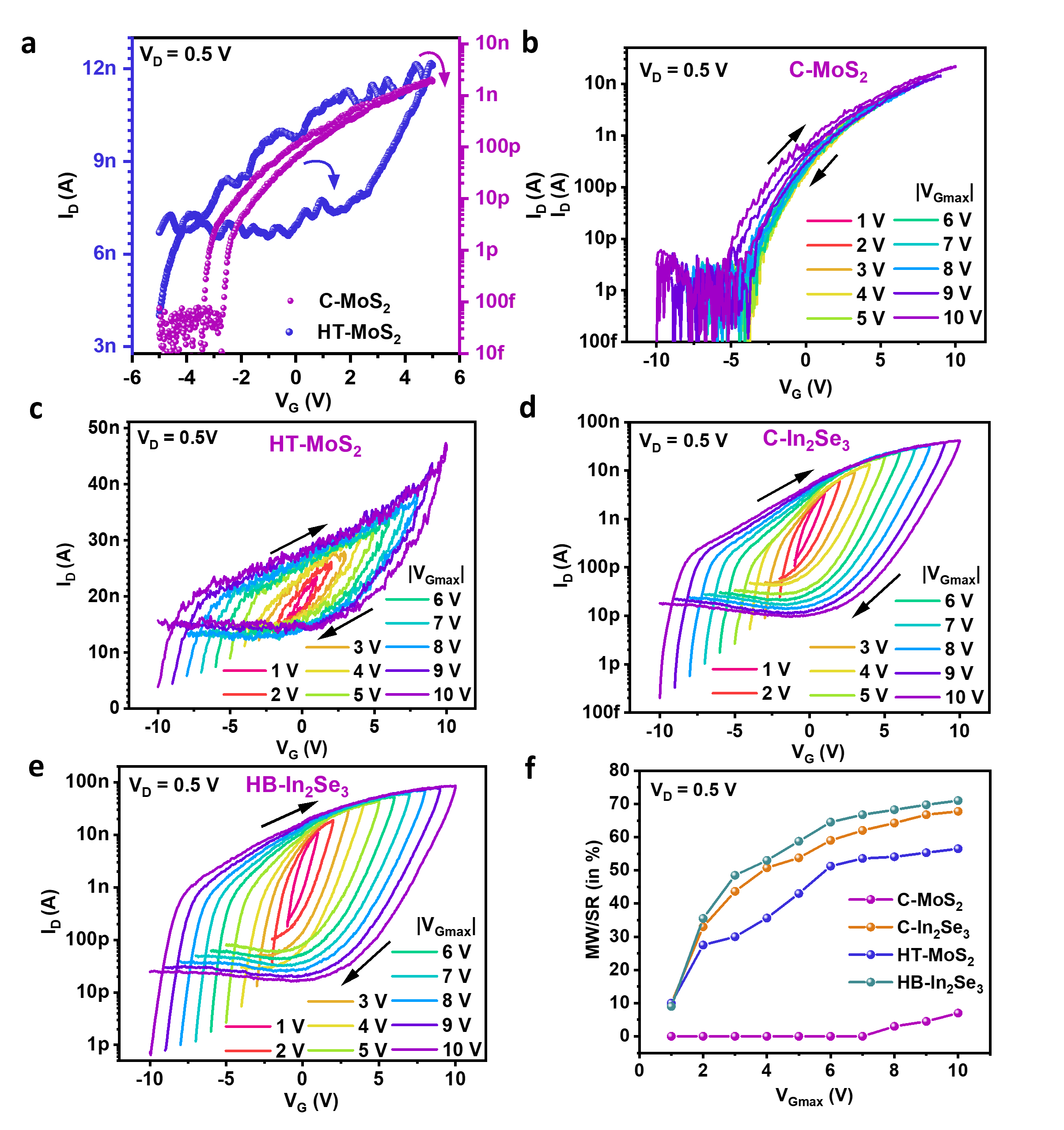}}
	\quad
\caption{(a) Comparison of $I_D$ - $V_G$ characteristics between C-MoS$_2$ and HT-MoS$_2$ FETs at $V_D$ = 0.5 V.  $I_D$ - $V_G$ characteristics of (b) C-MoS$_2$, (c) HT-MoS$_2$, (d) C-In$_2$Se$_3$, and (e) HB-In$_2$Se$_3$ FETs for different maximum gate bias values at $V_D$ = 0.5 V. (f) MW/SR ratios extracted as a function of gate bias sweep range from (b-e) at $V_D$ = 0.5 V. }
\label{s3}
\end{figure}

%%%%%%%%%%%%Supporting figure
\newpage
\textbf{Supporting Information 4:} \textbf{Output ($I_D$ - $V_D$) characteristics.}
\begin{figure}[H]	
{\includegraphics[width=1\textwidth]{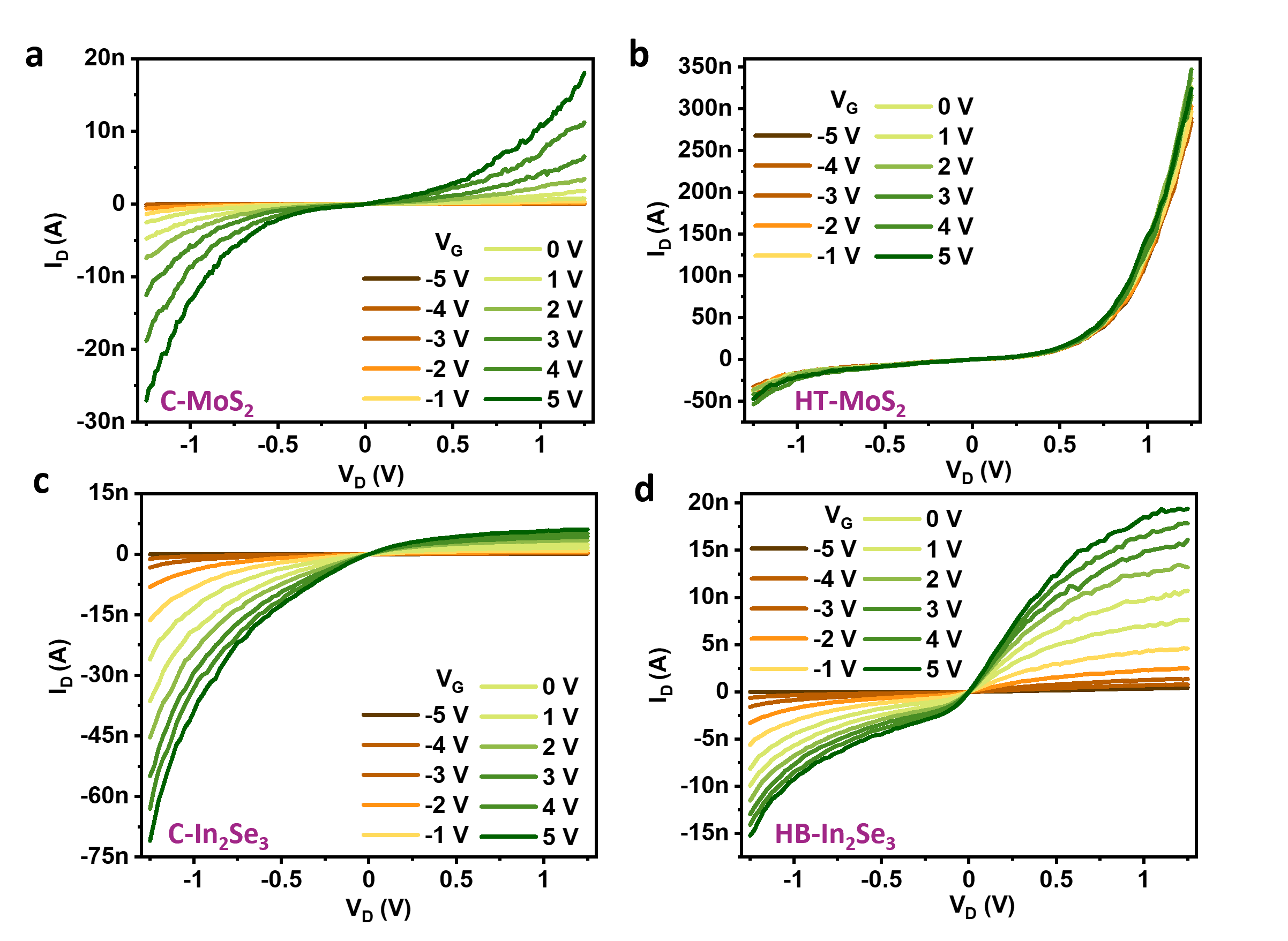}}
	\quad
\caption{Output ($I_D$ - $V_D$) characteristics of (a) C-MoS$_2$, (b) HT-MoS$_2$, (c) C-In$_2$Se$_3$, and, (d) HB-In$_2$Se$_3$ FETs for different gate bias values.}
\label{s4}
\end{figure}

%%%%%%%%%%%%Supporting figure5
\newpage
\textbf{Supporting Information 5:} \textbf{Electron energy band diagrams.}
\begin{figure}[H]	
{\includegraphics[width=0.9\textwidth]{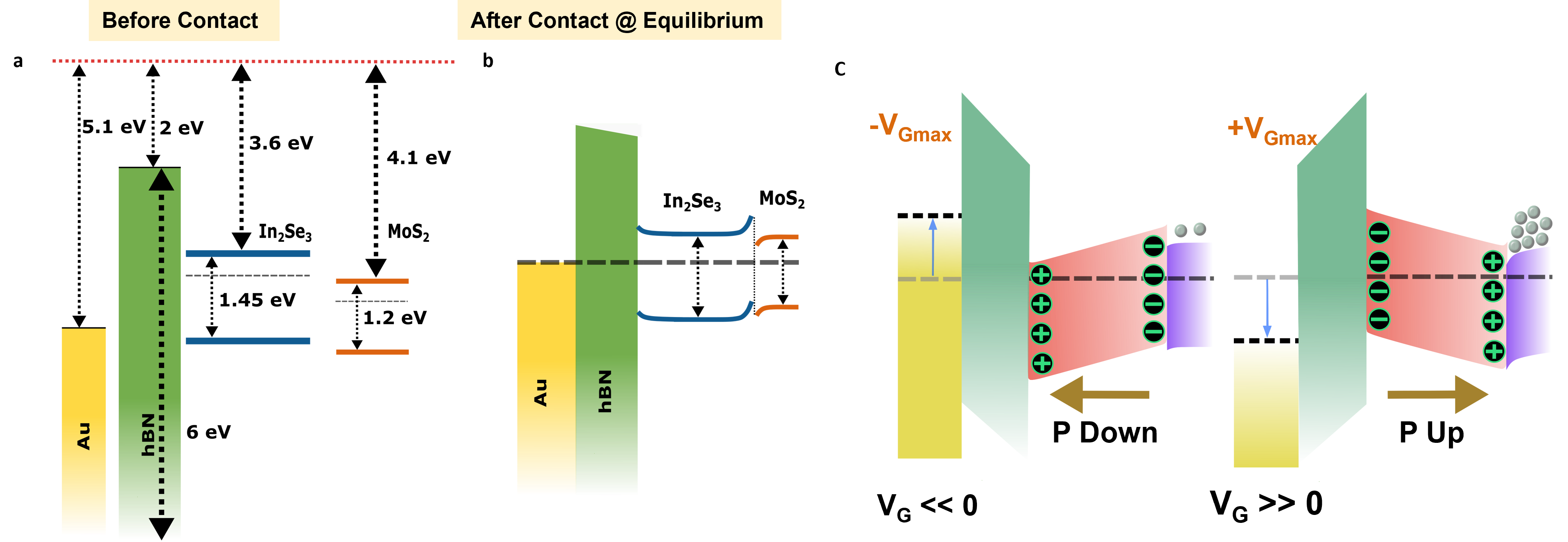}}
	\quad
\caption{Electron energy band diagrams along Au/hBN//In$_2$Se$_3$/MoS$_2$ stack (a) before contact, (b) after contact at equilibrium, and, (c) at -$V_{Gmax}$ (+$V_{Gmax}$) indicating high negative (positive) gate field-induced strong downward (upward) In$_2$Se$_3$ polarization.}
\label{s5}
\end{figure}

%%%%%%%%%%%%Supporting figure6
\newpage
\textbf{Supporting Information 6:} \textbf{Homosynaptic characteristics of HT-MoS$_2$ FET.}
\begin{figure}[H]	
{\includegraphics[width=1\textwidth]{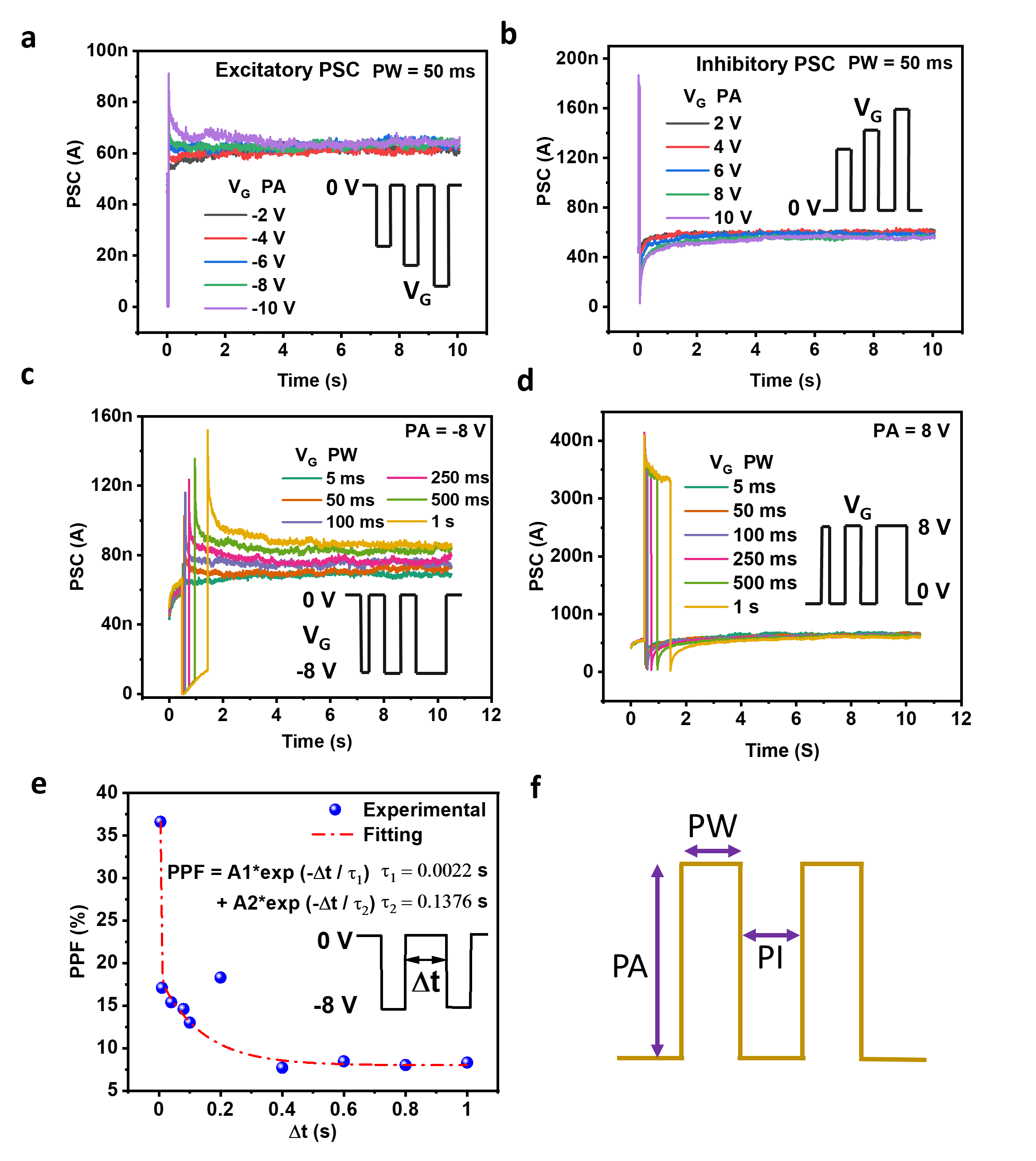}}
	\quad
\caption{The postsynaptic current of HT-MoS$_2$ FET versus time for a single gate pulse of different pulse amplitudes (PAs): (a) -$V_G$, (b) +$V_G$ and different pulse widths (PWs) at (c) $V_G$ = -8 V, and (d) $V_G$ = 8 V at $V_D$ = 1 V. (e)  Paired pulse facilitation (PPF) versus pulse intervals (PIs), $\Delta$t, for pulse amplitude (PA) of $V_G$ = -8 V and pulse width (PW) = 5 ms at $V_D$ = 1 V. (f) Definition of PA, PW, and PI.}

\label{s6}
\end{figure}

%%%%%%%%%%%%Supporting figure7
\newpage
\textbf{Supporting Information 7:} \textbf{P/D nonlinearity factor calculation.}

The equations used for the nonlinearity factor calculations are as follows:
%G$_P$ = G$_{min}$ +((G$_{max}$- G$_{min}$)*(1-exp(-n/A$_P$)))*(1/(1-exp(-1/A$_P$))), and 

%G$_D$ = G$_{max}$ - ((G$_{max}$- G$_{min}$)*(1-exp((1-n)/A$_D$)))*(1/(1-exp(1/A$_D$))), 
\begin{equation}
G_P = G_{\text{min}} + \left( (G_{\text{max}} - G_{\text{min}}) \cdot \left(1 - \exp\left(\frac{-n}{A_P}\right)\right) \right) \cdot \left(\frac{1}{1-\exp\left(\frac{-1}{A_P}\right)}\right)
\end{equation}
\begin{equation}
G_D = G_{\text{max}} - \left( (G_{\text{max}} - G_{\text{min}}) \cdot \left(1 - \exp\left(\frac{1-n}{A_D}\right)\right) \right) \cdot \left(\frac{1}{1-\exp\left(\frac{1}{A_D}\right)}\right)
\end{equation}

where n = pulse number/n$_{max}$, n$_{max}$ = N-1, N is the total number of pulses, and G$_{max}$ and G$_{min}$ denote the normalized maximum and minimum conductance levels. The nonlinearity factors, $\alpha$$_p$ and $\alpha$$_d$, are calculated by mapping A$_{P}$ and A$_{D}$ using DNN+Neurosim document.\cite{NeuroSim2024}

%%%%%%%%%%%%Supporting figure8
\newpage
\textbf{Supporting Information 8:} \textbf{Additional P/D measurements of homosynaptic HT-MoS$_2$ FET.}
\setcounter{figure}{7}  % Set the figure counter to 7
\begin{figure}[H]	
{\includegraphics[width=1\textwidth]{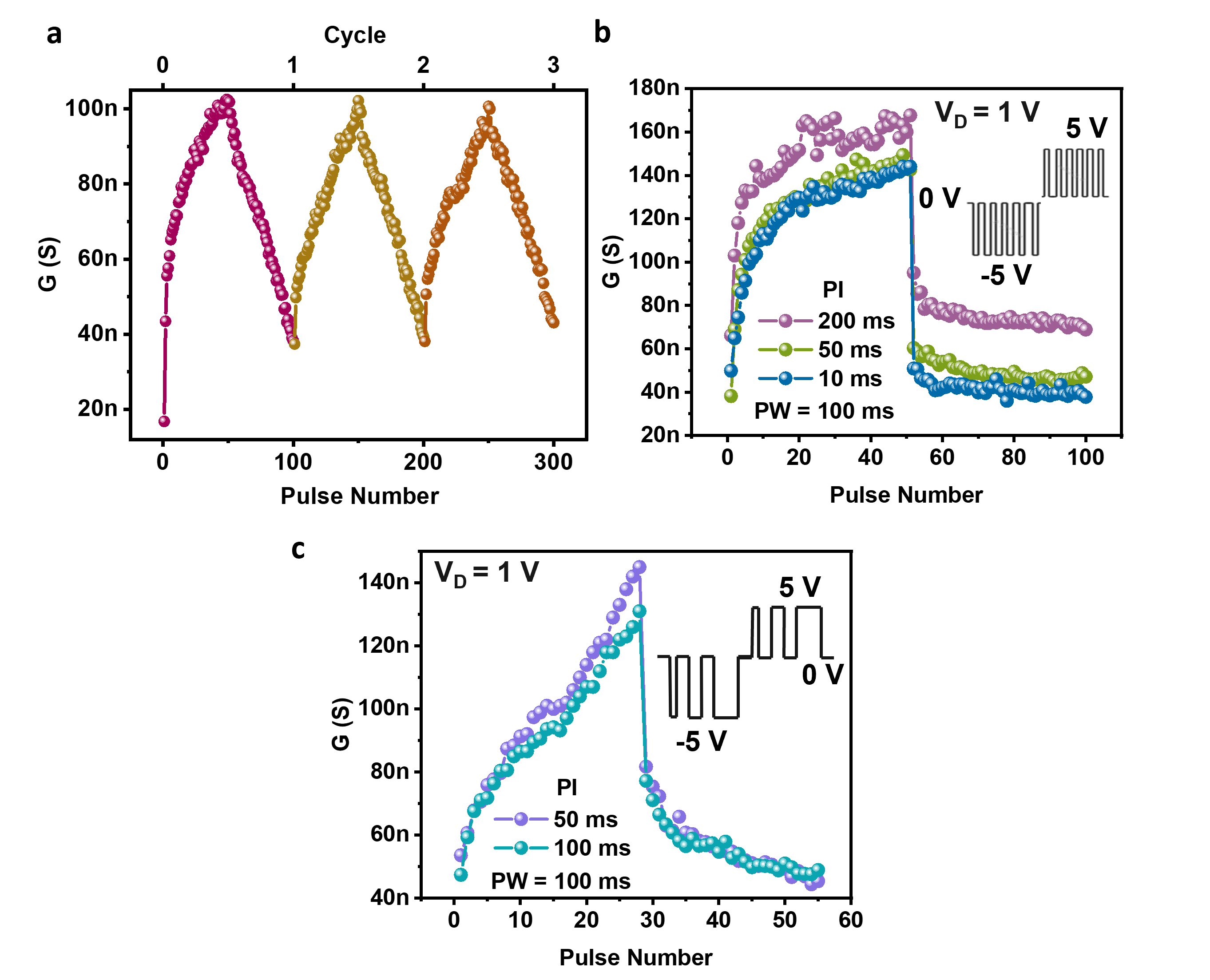}}
	\quad
\caption{Potentiation/depression (P/D) curves of HT-MoS$_2$ FET under different pulse schemes. P/D curves vs pulse number for (a) multiple cycles of incremental amplitude pulse scheme, (b) identical pulse scheme at a fixed PW = 100 ms for different PIs, and (c) varied pulse-width scheme for different PIs at a drain voltage of 1 V.}

\label{s8}
\end{figure}

\begin{table}
    \centering
    \begin{tabular}{|c|c|c|} \hline 
         Pulse Scheme&  $\alpha$$_p$&  $\alpha$$_d$\\ \hline 
         Identical&  5.65&  -\\ \hline 
         Incremental Amplitude&  1.74&  -0.07\\ \hline 
         Varied Pulse-width&  0.74&  -7.65\\ \hline
    \end{tabular}
    \caption{Nonlinearity factors evaluated from P/D curve versus pulse number of HT-MoS$_2$ FET for different pulse schemes.}
    \label{t1}
\end{table}

%%%%%%%%%%%%Supporting figure9
\newpage
\textbf{Supporting Information 9:} \textbf{Test accuracy for MNIST dataset on-chip inference (HT-MoS$_2$ synapse).}
\begin{table}
    \centering
    \begin{tabular}{|c|c|} \hline 
         Pulse Scheme&  Test Accuracy (\%)\\ \hline 
         Identical
&  97.08\\ \hline 
         Incremental Amplitude
&  97.21\\ \hline 
         Varied Pulse-width&  97.2\\ \hline
    \end{tabular}
    \caption{Test accuracy values evaluated using MNIST dataset for on-chip inference utilizing a fully connected neural network (FCNN) for HT-MoS$_2$ FET.}
    \label{t2}
\end{table}

\textbf{Supporting Information 10:} \textbf{STDP data modeling (HT-MoS$_2$ synapse).}

The exponential fitting equations used for the STDP curves are given by:

Y = Y$_0$ + A1*exp(-$\Delta$t/ $\tau$) for $\Delta$t>0

Y = -Y$_0$ - A1*exp(-$\Delta$t/ (-$\tau$)) for $\Delta$t<0

Y$_0$ denotes the residue, but with longer $\Delta$t, Y$_0$ is considered to be 0.

\newpage
\textbf{Supporting Information 11:} \textbf{Sensorimotor NN of sea mollusk (Aplysia).}

 Aplysia is a tiny sea mollusk with simple neural network containing only 10000 neurons.\cite{moroz2011aplysia} The gill withdrawal reflex (GWR) action in Aplysia is regulated by its sensorimotor neural network. This behavioral response can be extended to broader contexts, such as understanding learning and memory in human neural networks and creating therapeutic strategies for neurological disorders. 

GWR action in the mollusk gets triggered when the tail or siphon (which has sensory neurons (SN)) gets stimulated by an external event which in turn triggers the modulatory interneuron (MIN). In response to the external stimulus, MIN releases serotonin, which triggers synaptic plasticity in the motor neurons (MN), leading to retraction/withdrawal of the gill.\cite{pinsker1970habituation, chitwood2001serotonin}

\bibliography{supporting}